\documentclass[letterpaper]{article} 
\usepackage{aaai25}  
\usepackage{times}  
\usepackage{helvet}  
\usepackage{courier}  
\usepackage[hyphens]{url}  
\usepackage{graphicx} 
\usepackage{natbib}  
\usepackage{caption} 
\usepackage{algorithm}
\usepackage{algorithmic}
\usepackage{multirow}
\usepackage{subfigure}
\usepackage{amsmath}
\usepackage{enumitem}
\usepackage{amssymb} 
\usepackage{wasysym}
\usepackage{amssymb}
\usepackage{color}
\usepackage[most]{tcolorbox}
\usepackage{needspace}
\usepackage{newfloat}
\usepackage{listings}
\DeclareCaptionStyle{ruled}{labelfont=normalfont,labelsep=colon,strut=off} 
\floatstyle{ruled}
\newfloat{listing}{tb}{lst}{}
\floatname{listing}{Listing}
\title{BotSim: LLM-Powered Malicious Social Botnet Simulation}
\author{
    Boyu Qiao\textsuperscript{\rm 1,\rm2}, Kun Li\textsuperscript{\rm 1}\thanks{Corresponding Author.}, Wei Zhou\textsuperscript{\rm 1}, Shilong Li\textsuperscript{\rm 1,\rm2}, Qianqian Lu\textsuperscript{\rm 1}, Songlin Hu\textsuperscript{\rm 1,\rm2}
}
\affiliations{
    \textsuperscript{\rm 1}Institute of Information Engineering, Chinese Academy of Sciences\\
    \textsuperscript{\rm 2}School of Cyber Security, University of Chinese Academy of Sciences \\
    \{qiaoboyu, likun2, zhouwei, lishilong, luqianqian, husonglin\}@iie.ac.cn
}

\usepackage{bibentry}

\begin{document}

\maketitle

\begin{abstract}
Social media platforms like X(Twitter) and Reddit are vital to global communication. However, advancements in Large Language Model (LLM) technology give rise to social media bots with unprecedented intelligence. These bots adeptly simulate human profiles, conversations, and interactions, disseminating large amounts of false information and posing significant challenges to platform regulation. To better understand and counter these threats, we innovatively design BotSim, a malicious social botnet simulation powered by LLM. BotSim mimics the information dissemination patterns of real-world social networks, creating a virtual environment composed of intelligent agent bots and real human users. In the temporal simulation constructed by BotSim, these advanced agent bots autonomously engage in social interactions such as posting and commenting, effectively modeling scenarios of information flow and user interaction. Building on the BotSim framework, we construct a highly human-like, LLM-driven bot dataset called BotSim-24 and benchmark multiple bot detection strategies against it. The experimental results indicate that detection methods effective on traditional bot datasets perform worse on BotSim-24, highlighting the urgent need for new detection strategies to address the cybersecurity threats posed by these advanced bots. 
\end{abstract}

%
\begin{links}
    \link{Code}{https://github.com/QQQQQQBY/BotSim}
\end{links}

\section{Introduction}

In the modern digital era, online social networks (OSNs) such as X (formerly Twitter), and Reddit have become essential mediums for shaping human interaction due to their extensive connectivity and real-time information exchange. However, the prevalence of bots on these platforms poses a significant threat to OSN security \cite{cresci2020decade, ferrara2023social}. For example, social bots have played notable roles in major events like presidential elections  \cite{guglielmi2020next,pacheco2024bots} and global pandemics \cite{gallotti2020assessing,himelein2021bots}, where they disseminate misinformation and sway public opinion. Previous instances of social bots primarily stem from rule-based programs, however, recent advancements have integrated large language models (LLMs) that endow bots with more sophisticated, human-like capabilities \cite{yang2024anatomy}. This development has further intensified the problem of information pollution on OSNs \cite{sun2024exploring}. Therefore, upgrading current detection systems and understanding the characteristics of LLM-driven bots has become a critical priority.

Previous research methods have predominantly been developed using traditional bot datasets. For instance, Yang \textit{et al.} \shortcite{yang2020scalable} proposed a method that exploits differences in user profiles, while Cresci \textit{et al.} \shortcite{cresci2016dna} suggested identifying the longest common subsequence of user actions. With advancements in deep learning, new methods have emerged focusing on text semantic content and user interaction networks. Wei \textit{et al.} \shortcite{wei2019twitter} introduced the use of recurrent neural networks (RNNs) to encode posts and detect bots based on their semantic content. More recent methods, such as RGT \cite{feng2022twibot}, and BECE \cite{qiao2024dispelling} have employed graph neural networks (GNNs) and graph-enhanced strategies to improve detection performance. However, LLM-powered bots exhibit greater logical coherence and human-like qualities in profiles, text content, and interaction strategies, posing significant challenges to these existing detection methods \cite{feng2024does,ferrara2023social}. Therefore, collecting datasets of LLM-driven bots is essential for developing new detection techniques \cite{yang2024anatomy}. Traditional dataset collection methods, however, encounter the following two major challenges: 

\noindent \textbf{(1) Intelligent Challenges and Decline in Labeling Quality: } The intelligence of LLM-driven bots has significantly advanced, making manual annotation tasks much more challenging and leading to a notable decline in annotation quality \cite{zhang2024toward}. For instance, crowdsourcing tests conducted by Cresci \textit{et al.} \shortcite{cresci2017paradigm} revealed that manual annotators had an accuracy rate of less than 24\% when labeling social spam bots. Consequently, manual annotation has become unreliable, impairing the ability of detection models to differentiate between bots and genuine users. 

\noindent \textbf{(2) Ethical Constraints:} For ethical reasons, large-scale deployment of social bots disguised as humans in real social networks to obtain genuine annotations for research is subject to strict restrictions. This situation makes research more complex and challenging.

To address these challenges, we design a scalable malicious social botnet simulation framework called BotSim, upon which we construct an accurately labeled, LLM-driven bot dataset named BotSim-24. This dataset includes both real human accounts and LLM-driven agent bot accounts. To enhance the dataset's complexity, we implement a series of disguise techniques based on detection methods proposed in previous research focusing on bot profiles \cite{yang2020scalable}, textual content \cite{qiao2023social}, and interaction behavior patterns \cite{li2023multi}. By leveraging LLMs to analyze and simulate characteristics of real users, we construct a comprehensively disguised and highly human-like LLM-driven bot dataset to expose and challenge the limitations and weaknesses of existing detection methods. We then benchmark multiple bot detection strategies on the BotSim-24 dataset. The experimental results validate the effectiveness of the dataset and underscore the significant threat that advanced bots pose to network security.

Our contributions can be summarized as follows:

\begin{itemize}
    \item \textbf{BotSim Framework:} We are the first to propose a scalable LLM-driven malicious social botnet simulation framework, BotSim. This environment enables researchers to continuously track the latest bot evolution strategies and generate up-to-date datasets, thereby advancing the development of new detection methods. 

    \item \textbf{LLM-Driven Bot Dataset:} Leveraging the BotSim simulation framework, we meticulously construct a bot detection dataset based on interaction scenarios from Reddit. This dataset incorporates real Reddit users and LLM-driven bot accounts, providing a comprehensive range of interaction data that enhances existing resources for social bot detection research.

    \item \textbf{Experimental Evaluation:} We conduct extensive experiments on the BotSim-24 dataset to evaluate the performance of various social bot detection models. The results show that detection methods effective on traditional bot datasets perform poorly on BotSim-24, highlighting the urgent need for new detection strategies to address the cybersecurity threats posed by these advanced bots.

\end{itemize}

\begin{figure*}[ht]
\centering
\includegraphics[width=0.94\textwidth]{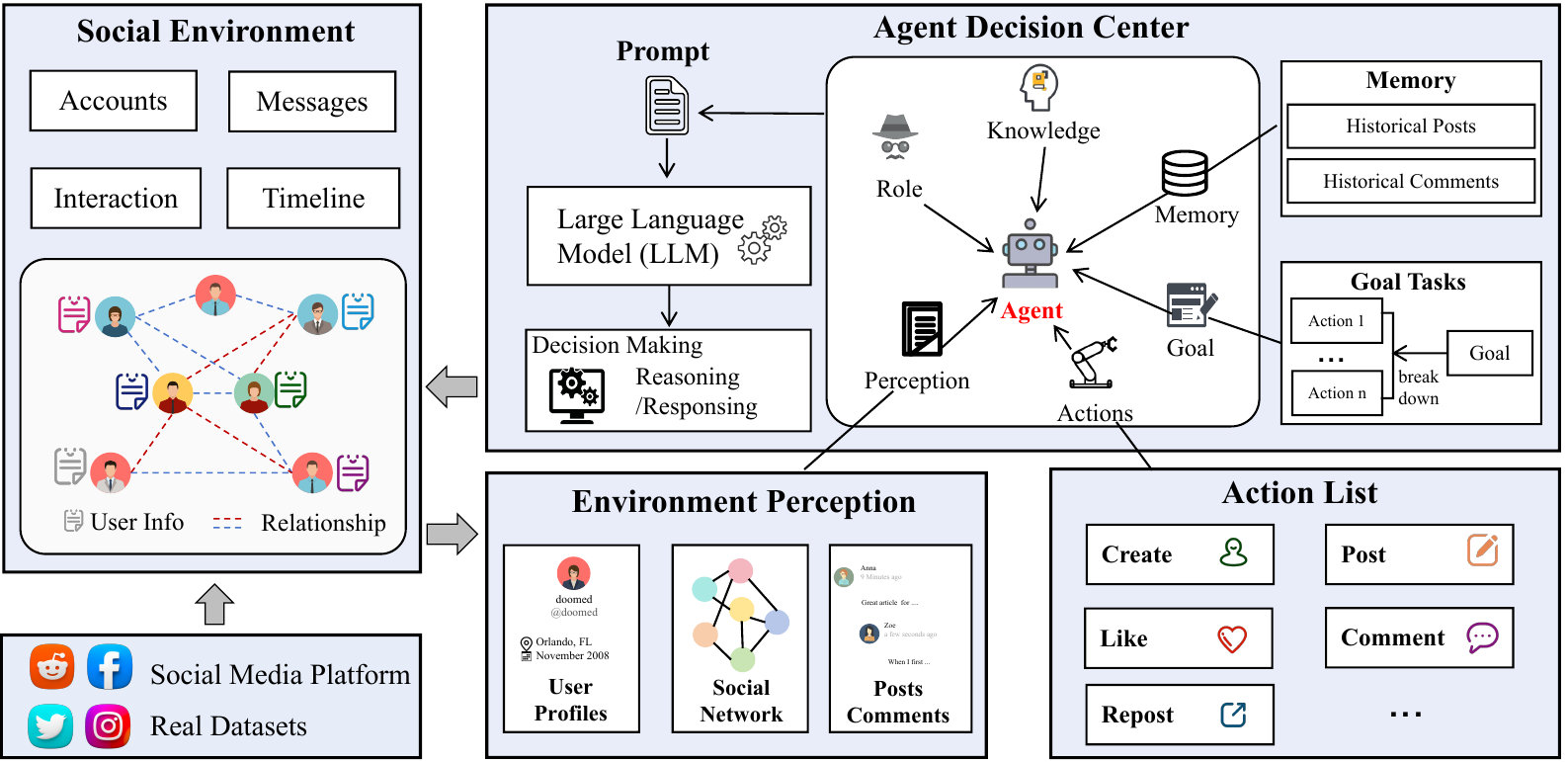}
\centering
\caption{The overall framework of  \textbf{BotSim}.}
\label{fig:1}
\end{figure*}

\section{BotSim: Botnet Simulation Framework}
The overall framework of BotSim is shown in Figure \ref{fig:1}, and it aims to model the activity characteristics and behavior patterns of LLM-driven malicious social bots in OSNs. BotSim consists of four components: the social environment, environmental perception, action list, and agent decision center.

\subsection{Preliminaries}
In this paper, we aim to use a botnet simulation framework to model the activity characteristics and behavior patterns of LLM-driven malicious bots on OSNs. The BotSim framework includes two types of users: human accounts from real social ecosystems, denoted as $U_H = \{{U_{h_1}}, {U_{h_2}}, ..., {U_{h_n}}\}$ and LLM-driven agent bot accounts, denoted as $U_B = \{U_{b_1}, U_{b_2}, ..., U_{b_m}\}$, where $n$ and $m$ represent the number of humans and bots, respectively. To simulate the continuous passage of time and the dynamic changes in interaction timing in real OSNs, we set up a timeline mechanism $T = \{t_1, t_2, ..., t_n\}$. In the timeline process, the set of interactions between users is represented as $D = \{U_B, U_H, E, T\}$ with $E = \{e_1, e_2, ..., e_n\}$ denoting the set of interaction relationships among users.

\subsection{Social Environment}
The social environment of BotSim is built from real social media ecosystem data and consists of account collection, message feeding, timeline setup, and interaction mode.


\noindent \textbf{Account Collection} \hspace{0.15cm} The account collection includes real human accounts $U_H$ and virtual Agent bot accounts $U_B$. Human accounts are sourced from data collected in real social environments, while the configuration and behavior of agent bot accounts are constructed by LLM-driven agents.


\noindent \textbf{Message Feeding} \hspace{0.15cm} Message feeding utilizes a dual-filtering mechanism based on timelines and recommendation functions. Initially, the message flow is filtered through the timeline, and then it is optimally ranked by the recommendation function to produce the final message stream.


\noindent \textbf{Timeline Setup} \hspace{0.15cm} The timeline setup $T = \{t_1, t_2, ..., t_n\}$ ensures the environment operates according to a predefined timeline logic. Additionally, each agent bot has its dedicated timeline, which is determined by the bot's activities and interactions with other accounts to meet the need for rapid simulation of long-time-span interactions.


\noindent \textbf{Interaction Mode} \hspace{0.15cm} The interaction patterns $E = \{e_1, e_2, ..., e_n\}$ must adhere to the interaction settings defined by the specific social media platform. Interactions between accounts are accompanied by message flow outputs, such as likes and comments on current messages.

\subsection{Environment Perception}
The environment perception mechanism is important in the operation of BotSim, which helps the agent to capture the dynamic changes of the social environment and accurately transfer the perceived multi-dimensional information to the agent decision center so that the agent can make adaptive decisions based on the environmental information.

In BotSim, account profiles, message stream updates, and complex interaction data collectively form the core elements of the social environment. To enhance the agents' understanding and responsiveness to these complex environments, we have designed clear and structured prompts to assist the LLM in comprehending environmental information. Detailed prompts can be found in Appendix B.1.

\subsection{Action List}

The action list integrates commonly used information dissemination interactions on social media, including the following actions: (1) \textbf{Create User:} Create a new user profile. (2) \textbf{Post:} Generate and publish original content based on background knowledge and preferences. (3) \textbf{Comment:} Reply to selected posts or comments. (4) \textbf{Repost:} Share posts to achieve targeted information dissemination. (5) \textbf{Like}: Like posts to enhance positive feedback during interactions. (6) \textbf{Browse:} Continue browsing the message stream based on the internal timeline if no preferred content is found. (7) \textbf{End}: Complete the mission and terminate the action.

BotSim provides a list of commonly used actions for information dissemination across various OSNs. Future research can select the appropriate actions based on specific needs and add new actions as required. Detailed description of the action list in Appendix B.2. 

\subsection{Agent Decision Center}

The Agent Decision Center, as the core component of BotSim, integrates multidimensional information including goal tasks, role settings, background knowledge, environmental perception, action lists, and memory data. Its primary function is to accurately plan and execute action decisions, driving the comprehensive operation of BotSim.

\noindent \textbf{Goal Tasks} \hspace{0.15cm} Goal tasks $G$ define the specific needs for information dissemination and guide the agent's actions. The operators set these goals, and then the LLM decomposes the goal tasks into manageable and planned actions $PA = \{pa_1, pa_2, ..., pa_k\}$ to ensure the goals are achieved. Prompts for goal tasks are detailed in Appendix B.3.

\noindent \textbf{Role Setting} \hspace{0.15cm} Role settings are crucial for the agent's decision-making process and include multidimensional attributes such as age, name, gender, preferences, education level, description, and geographic location. These attributes are applied to the profiles of created user accounts to help the agent establish a persona, enhancing both emotional expression and decision-making accuracy. More detailed information on role setup is provided in Appendix B.2.

\noindent \textbf{Background Knowledge} \hspace{0.15cm} Given that LLMs may struggle to capture new social dynamics and knowledge, providing background knowledge $KL$ can help LLMs generate relevant and novel content that aligns with goal tasks.

\noindent \textbf{Memory Mechanism} \hspace{0.15cm} The memory mechanism filters relevant posts and comments related to the current task from the agent's historical records. This mechanism assists the agent in responding appropriately. An example of memory information is presented in Appendix B.4.

\subsection{BotSim Execution Process}
The overall execution process of the agent bots in BotSim involves the following steps:
(1) \textbf{Specify the Platform: }Identify the social media platform to be simulated and gather the relevant data, including user profiles, messages, timestamps, and interaction data.
(2) \textbf{Define Goal Tasks: }Clearly outline the goal tasks and compile the necessary background knowledge.
(3) \textbf{Break Down Tasks: }Decompose the goal tasks into a series of executable actions, as detailed in Appendix B.3.
(4) \textbf{Formulate Environment Prompts: }Perceive changes in the simulated social environment and create appropriate prompts, as detailed in Appendix B.1.
(5) \textbf{Retrieve Memory Data: }Access historical posts and comments relevant to the goal task.
(6) \textbf{Construct and Execute Prompts: }Build prompts using environmental perception information, memory data, planned action sequences, role settings, and background knowledge. Use these prompts to instruct the LLM, which will return the required action parameters.
(7) \textbf{Update and Monitor: }Refresh the social environment and track the progress of the action sequence. If not completed, return to step (4). If completed, proceed to step (8). 
(8) \textbf{End: }Conclude the execution.

 A complete prompt example is provided in Appendix B.4, and the algorithm for this execution process is further explained in Appendix B.5.

\section{BotSim-24: LLM-driven Bot Detection Dataset}

In this section, we present BotSim-24, a bot detection dataset powered by LLM. Building on the BotSim framework, we simulate information dissemination and user interactions across six SubReddits on Reddit. This process results in the creation of the BotSim-24 dataset, which includes 1,907 human accounts and 1,000 LLM-driven agent bot accounts.

\subsection{Pre-Prepared Data}
We first introduce the real OSN data information that must be pre-prepared for the BotSim simulation.

\noindent \textbf{Reddit Social Environment Data Collection} \hspace{0.15cm} We choose six popular news-related SubReddits on Reddit to construct the social environment data for BotSim: ``worldnews'', ``politics'', ``news'', ``InternationalNews'', ``UpliftingNews'' and ``GlobalTalk''. We collect posts, first- and second-level comments, timestamps, and user profiles from these six SubReddits between June 20, 2023, and June 19, 2024. We filter and annotate the collected accounts, resulting in 1,907 human Reddit accounts. More detailed data filtering and statistical information are presented in Appendix C.1.

\noindent \textbf{Goal Tasks} \hspace{0.15cm} Our goal task is to create agent bots designed to spread disinformation within six news-oriented SubReddits. We focus on three highly debated international news events from 2023 to 2024: the ``Russia-Ukraine war,'' the ``Israeli-Palestinian conflict,'' and ``U.S. politics.'' Our objective is to disseminate disinformation related to these topics while concealing our activities by posting and engaging in discussions about a broad spectrum of international news on the SubReddits.

\noindent \textbf{Background Knowledge Collection} \hspace{0.15cm} To build the knowledge base for the three major news events and various international news used for our goal tasks, we collect real news from four authoritative international news sources —``BBC'', ``NBC News'', ``NYTimes'', and ``People's Daily'', as well as fact-checking sites ``Truthorfiction'' and ``Snopes''. The data spans from June 2023 to June 2024. This knowledge base helps the LLM generate content that is most relevant to the goal tasks. More detailed statistics are in Appendix C.2.

\noindent \textbf{User Role} \hspace{0.15cm} Role settings in BotSim are used to construct the profiles of agent bots. Usernames and descriptions are generated by LLM simulation cases, while age, gender, education level, and geographic location are randomly assigned based on weighted statistics from Reddit\footnote{https://explodingtopics.com/blog/reddit-users}. Additionally, since the goal tasks involve international news,  political ideology settings are included in the role settings \footnote{https://news.gallup.com/poll/388988/political-ideology-steady-conservatives-moderates-tie.aspx}. This information is intended to assist the agent Bots in interactions, but the BotSim-24 dataset only provides profile information relevant to Reddit.

\begin{table}[ht]
\centering
\setlength\tabcolsep{5pt} 
\renewcommand\arraystretch{1.2}
\begin{tabular}{|c|c|c|c|c|}
\hline
\textbf{SubReddit}         & \textbf{Posts} & \textbf{Users} & \textbf{1-Coms} & \textbf{2-Coms} \\ \hline
worldnews         & 14,626      & 1,405  & 15,740           & 859             \\ 
politics          & 2,4074      & 1,744  & 39,704           & 3,155             \\
news              & 8,465       & 1,471  & 11,685           & 441               \\
InternationalNews & 3,906       & 554    & 5,477            & 311           \\
UpliftingNews     & 1,219       & 266    & 1,148            & 35             \\
GlobalTalk        & 342         & 342    & 472              & 16               \\ \hline
Total             & 52,632      & 2,907  & 74,226          & 4,817           \\ \hline
\end{tabular}
\caption{Distribution of users, posts, and comments among six SubReddits. `1-Coms' means first-level comments, `2-Coms' means second-level comments. The total number of users is not the sum of users participating in different SubReddits, but the number of accounts participating in the social environment.}
\label{Table:1}
\end{table}

\subsection{Bot Data Construction}
Previous detection methods have primarily focused on identifying bot accounts that lack sufficient anthropomorphic features in areas such as profile metadata (Value or Boolean information) \cite{cresci2016dna, moghaddam2022friendship, beskow2018bot}, textual content \cite{qiao2023social, liu2023botmoe}, and interaction patterns \cite{feng2021botrgcn, peng2022domain}. Our goal is to create highly human-like bot accounts, driven by LLMs and based on the BotSim framework, to challenge these detection algorithms. To achieve this, the bots must effectively disguise themselves in these key areas to evade detection.

The disguise strategies we implement for bot accounts are as follows: (1) \textbf{Metadata Disguise:} We statistically analyze six types of value-type metadata from real Reddit users, including the number of posts, the number of first-level comments, the number of second-level comments, the ratio of posts to comments, posting frequency, and the number of active SubReddits. We then use LLM to integrate this statistical information to generate human-like metadata for bot accounts, effectively achieving metadata disguise. (2) \textbf{Textual Content Disguise:} The posts and comments of bot accounts are generated by LLMs based on contextual knowledge, user role information, browsing content, and other relevant factors. Unlike traditional bots, which often produce posts and comments with inconsistent contextual semantics, LLM-driven bots utilize advanced text understanding and generation capabilities to create contextually coherent and logically sound content, effectively disguising the textual output. (3) \textbf{Interaction Disguise:} On BotSim Reddit, interactions between accounts include first-level and second-level replies. The specific posts or comments that bot accounts reply to are autonomously determined by the LLM based on the goal task and browsed information. This method leverages the LLM's analytical capabilities, distinguishing it from previous rule-based settings, and thereby achieving interaction disguise. We present a more detailed data statistical analysis and the process of constructing bot data in Appendix C.3 and C.4.

After setting up the data information and construction strategies required for BotSim, we selected GPT4o-mini as the LLM for generating the BotSim-24 dataset. The BotSim-24 contains users’ profiles, post and comment information, and relationship information. We present the statistical information of the constructed BotSim-24 dataset in Table \ref{Table:1}. 

\begin{table*}[ht]
\centering
\renewcommand\arraystretch{1.1}
\setlength{\tabcolsep}{2mm}
\begin{tabular}{|c|c|c|c|c|c|c|c|c|c|}
\hline
\textbf{Dataset} & \textbf{Users} & \textbf{Human} & \textbf{Bot}   & \textbf{Training} & \textbf{Validation} & \textbf{Test} & \textbf{Edges}  &\textbf{Edge Types} & \textbf{Communities} \\ \hline
\textbf{BotSim-24}     & 2,907 & 1,907 & 1,000 & 2,304    & 582        & 291     & 46,518 & 3 & 6           \\ \hline
\end{tabular}
\caption{Statistics of our BotSim-24 dataset.}
\label{Table:2}
\end{table*}

\begin{table*}[ht]
\small
\centering
\renewcommand\arraystretch{1.1}
\setlength{\tabcolsep}{0.5mm}
\begin{tabular}{|c|c|cccc|c|c|cc|ccc|}
\hline
\multirow{2}{*}{\textbf{Dataset}}   & \multirow{2}{*}{\textbf{Score}} & \multicolumn{4}{c|}{\textbf{Metadata-based}}                   & \textbf{Text-based} & \textbf{Meta-Text} & \multicolumn{2}{c|}{\textbf{Homo-GNN}} & \multicolumn{3}{c|}{\textbf{Heter-GNN}}           \\ \cline{3-13} 
                           &                        & \textbf{AB}          & \textbf{RF}          & \textbf{DT}          & \textbf{SVM}         & \textbf{Wei \textit{et al.}}        & \textbf{Roberta+NN}    & \textbf{GCN}           & \textbf{GAT}           & \textbf{BotRGCN}     & \textbf{RGT}         & \textbf{S-HGN}       \\ \hline
                \multicolumn{13}{|c|}{\textbf{Manual Labeling Strategy}}           \\ \hline
\multirow{2}{*}{Cresci-15} & Acc                    & 95.9±0.3    & 97.0±0.8    & 96.2±1.3    & 96.6±0.2    & 96.18±1.5  & 95.14±0.5   & 98.2±0.6      & 98.1±0.2      & \underline{98.5±0.4}    & \textbf{98.6±0.3}    & 97.5±0.5    \\
                           & F1                     & 95.5±0.3    & 96.7±0.9    & 95.9±1.4    & 96.3±0.3    & 82.65±2.2  & 96.19±0.4   & 98.0±0.4      & \underline{98.0±0.1}      & 97.3±0.5    & \textbf{98.5±0.2}    & 97.2±0.5    \\ \hline
\multirow{2}{*}{Cresci-17} & Acc                    & \underline{91.2±0.2}    & 89.1±0.2    & 86.2±0.2    & 84.1±0.3    & 89.30±0.3  & \textbf{96.22±0.4}   & /             & /             & /           & /           & /           \\
                           & F1                     & \underline{83.4±0.2}    & 80.9±0.2    & 76.4±0.2    & 72.8±0.3    & 78.40±0.2  & \textbf{97.37±0.4}   & /             & /             & /           & /           & /           \\ \hline
\multirow{2}{*}{Twibot-20} & Acc                    & 85.7±0.4    & 85.0±0.5    & 80.1±0.5    & 85.2±0.3    & 71.26±0.1  & 85.11±0.3   & 77.2±1.2      & 83.2±0.4      & \underline{86.8±0.5}    & \textbf{86.9±0.3}    & 85.4±0.3    \\
                           & F1                     & 85.6±0.4    & 84.9±0.5    & 80.0±0.5    & 84.8±0.4    & 75.33±0.1  & 87.02±0.2   & 76.6±0.4      & 81.9±0.5      & \underline{86.6±0.4}    & \textbf{86.7±0.4}    & 85.3±0.2    \\ \hline
\multirow{2}{*}{MGTAB-22}  & Acc                    & 90.1±0.9    & 89.5±0.4    & 87.1±0.5    & 88.7±1.4    & /          & 84.8±1.6   & 85.8±1.3      & 87.0±1.3      & 89.6±0.8    & \textbf{92.1±0.4}    & \underline{91.4±0.4}    \\
                           & F1                     & 87.7±1.1    & 86.8±0.5    & 83.7±0.7    & 85.3±1.7    & /          & 68.9±4.3   & 78.3±1.7      & 82.3±2.1      & 87.2±0.7    & \textbf{90.4±0.5}    & \underline{88.7±0.6}    \\ 
                           \hline
        \multicolumn{13}{|c|}{\textbf{Weak Labeling Strategy}}                   \\ \hline
\multirow{2}{*}{Twibot-22} & Acc                    & 69.3±0.5 & 74.3±0.7  & 72.6±0.8 & 76.4±0.9 & 70.2±0.1           & 72.6±4.0   & 78.3±1.3   & \underline{79.3±0.8}   & \textbf{79.6±0.4 } & 76.5±0.4 & 76.7±1.3 \\
                           & F1                     & 34.8±0.5 & 30.4±0.6 & 51.6±0.6 & 54.6±0.8 &  53.6±1.4    &  47.5±0.3    & 54.8±1.0   & \underline{55.6±1.1}   & \textbf{57.6±1.4}   & 43.1±0.5 & 45.7±0.5  \\ \hline
            \multicolumn{13}{|c|}{\textbf{Simulation Labeling Strategy }}               \\ \hline
\multirow{2}{*}{\textbf{BotSim-24}} & { Acc}                    & {77.5±3.2} & {75.7±2.2} & {71.4±2.1} & {74.4±2.2} & {50.8±2.9}           & {67.6±4.0}   & {72.7±2.2}   & {80.3±1.4}   & {\textbf{89.9±1.8}} & {82.3±2.2} & {\underline{87.7±1.3}} \\
                           & { F1}                     & {74.8±3.5} & {72.4±1.6} & {68.5±2.2} & {69.8±2.4} &  {50.4±1.4}          & {30.5±6.1}   & {50.5±5.5}   & {73.1±3.6}   & {\textbf{86.7±3.0}} & {76.4±3.1} & {\underline{83.1±2.9}} \\ \hline
\end{tabular}
\caption{Performance of the baseline method on 6 datasets. Each baseline is performed five times with different seeds and we report the average performance and standard deviation. The best and second-best results are highlighted in bold and underlined. ``/" indicates that the dataset does not contain support for the corresponding method. ``Homo-GNN'' indicates homogeneous GNNs and ``Heter-GNN'' indicates GNNs. We show the labeling strategy of different datasets.}
\label{Table:3}
\end{table*}

\subsection{Dataset Process}
In this section, we describe the construction of user features and relationships in the BotSim-24 dataset.

\noindent \textbf{User Features Construction} \hspace{0.15cm} Following the user feature processing methods used in the Cresci-15 \cite{cresci2015fame} and Twibot-20 \cite{feng2021twibot} datasets, we process the user features in BotSim-24 into metadata features and text features. Metadata features include Reddit user profile information, as detailed in Appendix C.1, and additional derived features based on basic profile information, totaling 10 standardized numerical data types. Text features include both posts and comments made by users. Compared to previous bot detection datasets that contain only user posting information, BotSim-24 also incorporates bi-level comment information.  

\noindent \textbf{User Relationships Construction} \hspace{0.15cm}Clarifying the types of interaction relationships between users is crucial for subsequent graph-based bot detection methods. We categorize user relationships into three types: first-level comment users and post users, second-level comment users and post users, and first-level comment users and second-level comment users. We record the number of comments exchanged between users, which can be used as edge weights in future graph structures to assist in detection. Appendix C.5 provides more detailed information on user features and user relationships, as well as comparisons with other datasets.

\section{Experiment}

\subsection{Experiment Settings}
\noindent \textbf{Parameter Settings} \hspace{0.15cm} Our experiments are conducted on four Tesla V100 GPUs with 32GB of memory. Detailed hyperparameter settings can be found in Appendix A.1.

\noindent \textbf{Baseline} \hspace{0.15cm} We evaluate various commonly used methods for bot detection on BotSim-24, including feature-based, text-based, and graph-based approaches. These methods encompass the the Adaboost classifier (AB) \cite{hastie2009multi}, decision tree (DT) \cite{lepping2018wiley}, random forest (RF) \cite{yang2020scalable}, support vector machine (SVM) \cite{boser1992training},  the approach proposed by Wei \textbf{et al.} \shortcite{wei2019twitter}, Roberta+NN, and both homogeneous graph methods (GCN \cite{kipf2016semi}, GAT \cite{velivckovic2017graph}) and heterogeneous graph approaches (S-HGN \cite{lv2021we}, BotRGCN \cite{feng2021botrgcn}, RGT \cite{feng2022twibot}). A more detailed description is provided in Appendix A.2.

\noindent \textbf{Datasets} \hspace{0.15cm} We evaluate BotSim-24 alongside five publicly available bot detection datasets: Cresci-15 \cite{cresci2015fame}, Cresci-17 \cite{cresci2017paradigm},  TwiBot-20 \cite{feng2021twibot}, TwiBot-22 \cite{feng2022twibot}, and MGTAB-22 \cite{shi2023mgtab}. Consistent with the division used in TwiBot-20 and MGTAB-22, we randomly divide all datasets into training, validation, and test sets with a ratio of 7:2:1. Table \ref{Table:2} shows the division of the BotSim-24 dataset. More detailed comparisons are presented in Appendix A.3.

\subsection{Experiment Results}

We evaluate the performance of 11 baseline methods across 6 datasets, with each baseline executed 5 times. The average performance and standard deviation are reported. The labeling strategy, detection accuracy, and F1-scores for each dataset are presented in Table \ref{Table:3}. Our key findings are summarized as follows:

\noindent \textbf{The BotSim-24 dataset presents greater challenges for baseline detection methods.} Table \ref{Table:3} indicates that the 11 baseline methods perform poorly on both the Twibot-22 and BotSim-24 datasets. The weak results on Twibot-22 are likely due to its reliance on low-quality weak supervision for labeling. Despite the high label reliability of BotSim-24, it still underperforms compared to other reliable datasets like Cresci-15 and Twibot-20, due to its effective camouflage of various features. Specifically, for metadata-based methods, the BotSim-24 dataset undermines the performance of traditional machine-learning approaches due to its successful camouflage of metadata features. For text-based methods, the exceptional text comprehension and generation capabilities of LLM make it nearly impossible to distinguish between bots and humans in such highly human-like content. Consequently, Wei \textit{et al.}'s method performs almost like random guessing. Furthermore, the ``Roberta + NN'' neural network method, which combines text and metadata features, not only fails to improve detection performance but also shows negative gains, highlighting the effectiveness of bots’ profiles and text disguises.

\noindent \textbf{Graph-based Methods Perform Better in Detecting LLM-driven Bots.} The detection performance after fusing relational edges is superior to that of methods only based on metadata and textual content. Although the rapid development of LLM technology has greatly enhanced the anthropomorphic nature of bot accounts, the complexity and dynamics of human-established relationships are still difficult to be fully modeled by LLM. This finding emphasizes the indispensability of inter-user relationship information in bot detection, and we believe it is a key clue for future research to distinguish human and machine behaviors.

\noindent \textbf{Methods Based on Heterogeneous Graphs Outperform Homogeneous Graphs.} The superior performance of heterogeneous graphs is primarily due to their ability to effectively utilize different types of edge relationships within the graph. This capability reveals diverse interaction patterns between users, allowing GNNs to gather more comprehensive information and enhance detection capabilities. Additionally, we observe that RGCN and S-HGN exhibit excellent performance on the BotSim-24 dataset. This is not only due to the excellence of RGCN and S-HGN design but also affected by the dataset. In the subsequent experimental analysis, we further elucidate the key factors contributing to their superior detection performance.

\subsection{Experimental Analysis}

\noindent \textbf{The Reason For the Excellent Performance of GNN Method.} In BotSim, human data is generated in timeline order, meaning that real human users do not participate in interactions. Consequently, the BotSim-24 dataset includes interactions between humans (human $\leftrightarrow$ human), bots and humans (bot $\rightarrow$ human), and bots (bot $\leftrightarrow$ bot), but it does not include interactions initiated by humans towards bots (human $\rightarrow$ bot). This creates an incomplete graph structure, missing directed edges from human to bot nodes, as detailed in Appendix A.4. This structural incompleteness enables GNNs to identify the intrinsic differences between human and bot accounts, which contributes to their superior detection performance.

\begin{figure}[htb]
    \centering
    \subfigure[RGCN Edge Perturbation]{
    \begin{minipage}[t]{0.22\textwidth}
        \centering
        \hspace{-0.2cm}
        \includegraphics[width=0.98\textwidth]{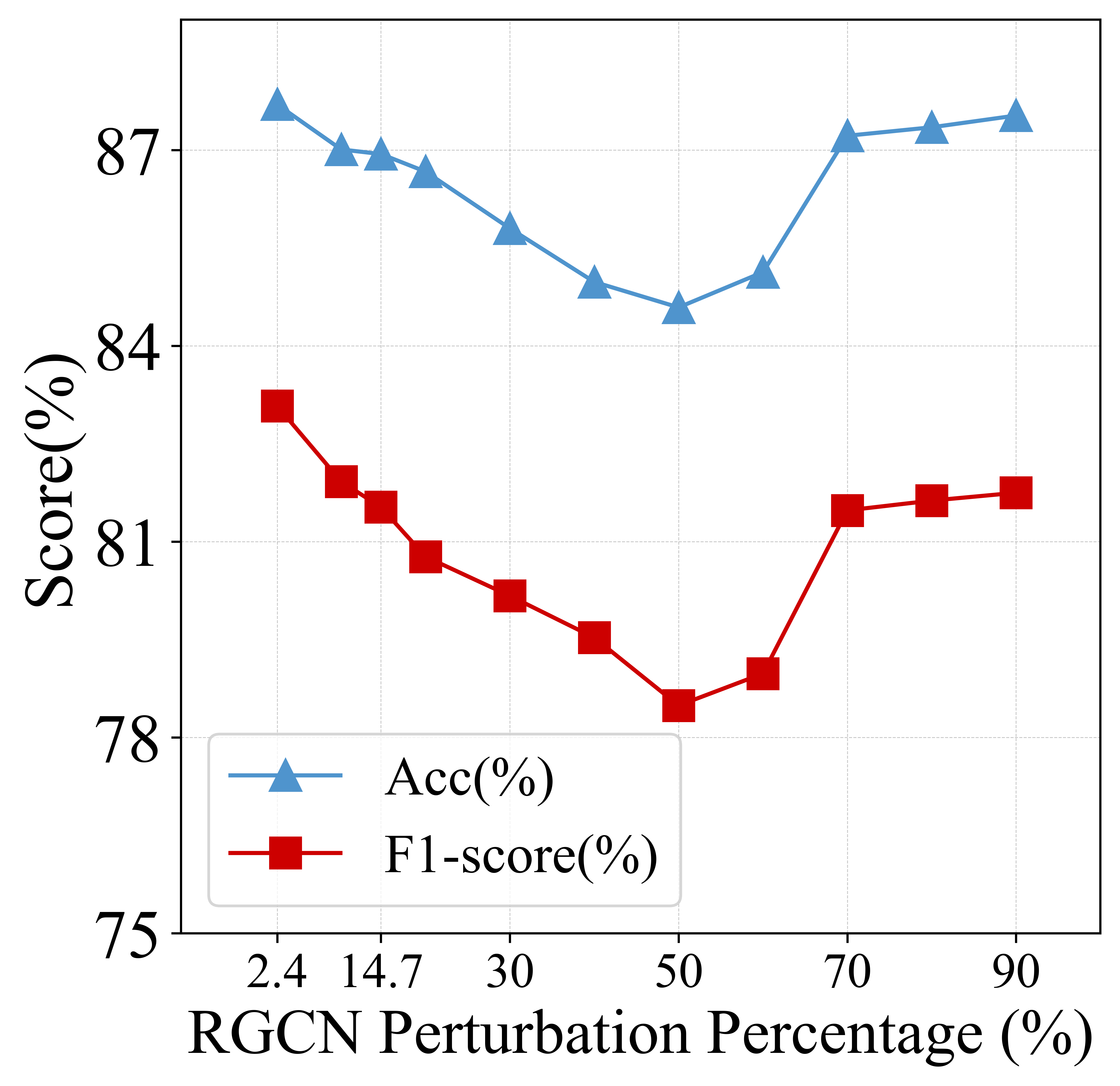}
    \end{minipage}
}%
\subfigure[S-HGN Edge Perturbation]{
    \begin{minipage}[t]{0.22\textwidth}
        \centering
        \hspace{-0.2cm}
        \includegraphics[width=0.98\textwidth]{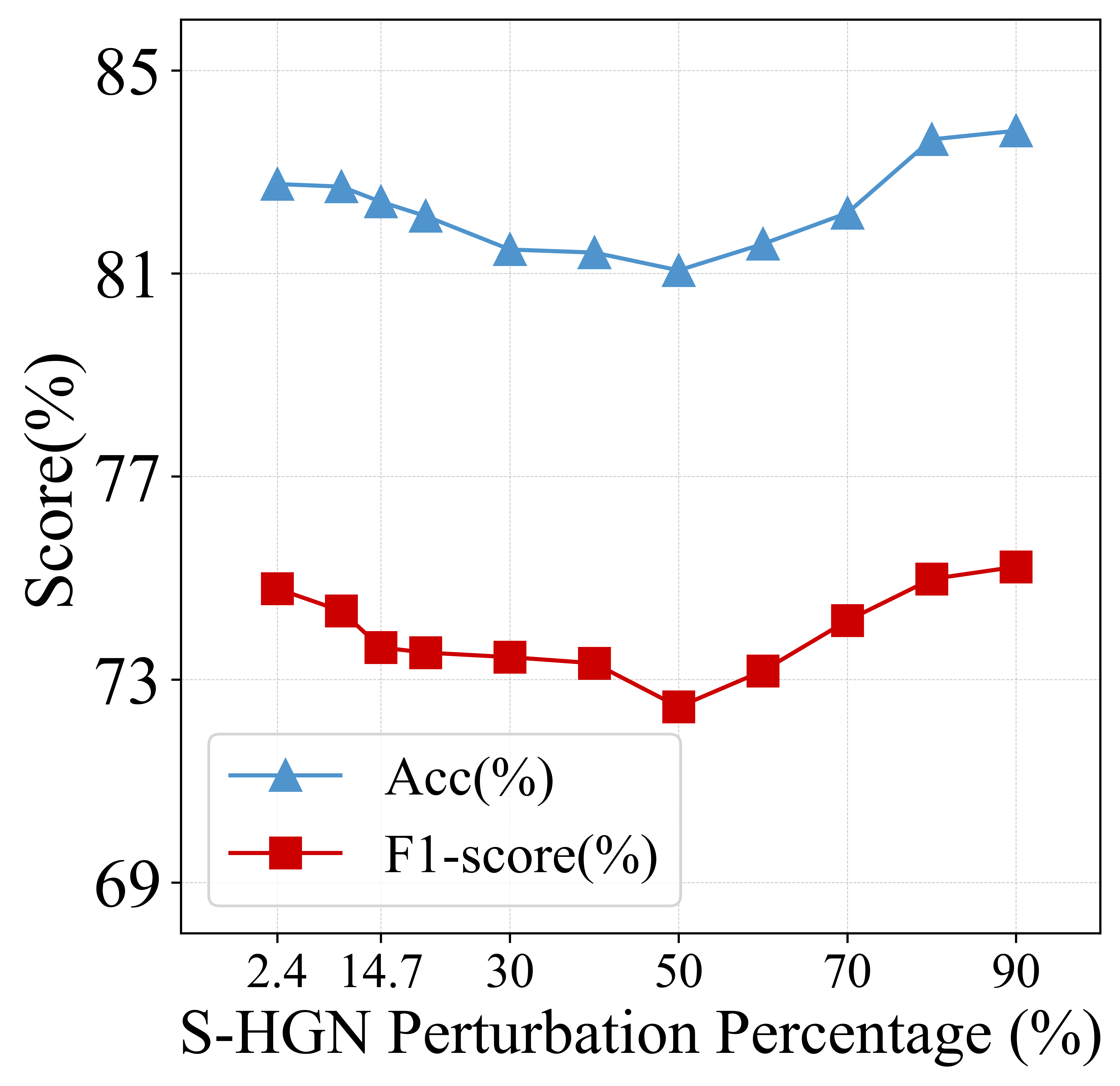}
    \end{minipage}
}%
    \caption{The impact of different proportions of edge perturbations on RGCN and S-HGN detection performance.}
    \label{fig:2}
\end{figure}

\noindent \textbf{Edge Perturbation Experiment.} To further validate our observations and hypotheses, we design an edge perturbation experiment. This experiment randomly reverses the direction of a proportion of the original edges to simulate varying levels of interaction between humans and bots. We count the interaction ratios between humans and bots in three real-world datasets in Table \ref{Table:4}: 0\%, 14.7\%, and 2.4\% for Cresci-15, TwiBot-20, and MGTAB-22, respectively. The differences in interaction ratios may be due to varying proportions of bots in different events. We then visualize edge perturbations using RGCN and S-HGN at these ratios and additional ones in Figure \ref{fig:2}. Results indicate that detection performance initially declines and then improves with increasing perturbation ratios. When the perturbation ratio reaches 50\%, performance drops but then rebounds. This is because introducing more directed edges from humans to bots allows the GNN to effectively capture these structural differences, enhancing detection performance.

\begin{table}[ht]
\centering
\renewcommand\arraystretch{1.1}
\setlength{\tabcolsep}{1.5mm}
\begin{tabular}{|c|c|c|c|}
\hline
\textbf{Dataset}                 & \textbf{Cresci-15} & \textbf{Twibot-20} & \textbf{MGTAB-22} \\ \hline
Human $\rightarrow$ Bot & 0         & 313          & 1140         \\ 
Edge Num          & 13,130          &  2,133         & 47,907         \\ \hline
\textbf{Proportion}              &  \textbf{0\%}         & \textbf{14.7\% }         &  \textbf{2.4\%}       \\ \hline
\end{tabular}
\caption{The interaction ratios between humans and bots. We randomly select the same number of humans and bots from these three datasets as in BotSim-24 and count the number of interaction edges between humans and bots. ``Human $\rightarrow$ Bot'' denotes the number of human-bot interaction edges and ``Edge Num'' denotes the total number of edges.}
\label{Table:4}
\end{table}

\noindent \textbf{Discussion.} \textbf{LLM-driven bots are becoming increasingly difficult to detect.} The BotSim-24 dataset does not include interactions between humans and bots. Statistics in Table \ref{Table:4} show that such interactions are also relatively sparse in actual OSNs. \textbf{However, as LLM-powered bots become more prevalent, their high human-like characteristics will inevitably lead to an increase in human-bot (human $\leftrightarrow$ bot) interactions. As demonstrated by our edge perturbation experiments, this trend will challenge and undermine the effectiveness of GNN-based methods.}  Furthermore, Table \ref{Table:5} offers a detailed overview of the performance of various LLMs in account detection tasks based on textual content. Additionally, Figure 4 in Appendix A.5 visually illustrates findings on the accuracy of human annotators. These results highlight the difficulty LLMs face distinguishing between text they generate and text authored by humans. Human annotators also struggle to achieve high accuracy in this regard. For additional details, please refer to Appendix A.5. This underscores the critical challenge of detecting LLM-driven bots and emphasizes the urgent need for innovative detection strategies to keep pace with their evolving capabilities.

\begin{table}[ht]
\centering
\renewcommand\arraystretch{1.1}
\setlength{\tabcolsep}{2.5mm}
\begin{tabular}{|c|c|c|}
\hline
\textbf{LLM}                                  & \multicolumn{1}{|c|}{\textbf{Acc(\%)}}   & \multicolumn{1}{|c|}{\textbf{F1-score(\%)}}    \\ \hline
\multicolumn{3}{|c|}{\textbf{LLAMA 3-8B}}  \\ \hline              
Zero-Shot Text                       & \multicolumn{1}{c}{54.82} & \multicolumn{1}{|c|}{54.22} \\  
2-Shot Text                          & \multicolumn{1}{c}{55.11} & \multicolumn{1}{|c|}{54.40} \\ 
5-Shot Text                          & \multicolumn{1}{c}{57.79} & \multicolumn{1}{|c|}{52.89} \\ \hline
\multicolumn{3}{|c|}{\textbf{ChatGLM 3-6B}}  \\ \hline            
Zero-Shot Text                       & 42.61                     & 9.72                      \\ 
2-Shot Text                          & 42.62                     & 43.94                     \\ 
5-Shot Text                          & 47.78                     & 54.22                     \\ \hline
\multicolumn{3}{|c|}{\textbf{GPT-3.5-turbo-ca}}       \\ \hline   
Zero-Shot Text                       & 42.96                     & 50.89                     \\ 
2-Shot Text                          & 48.80                     & 53.29                     \\ 
5-Shot Text                          & 53.61                     & 57.94                     \\ \hline
\multicolumn{3}{|c|}{\textbf{GPT-4-turbo-ca}}      \\ \hline               
Zero-Shot Text                       & 39.18                     & 22.03                     \\ 
2-Shot Text                          & 59.79                & 32.03                     \\ 
5-Shot Text                          & 70.76                    & 63.58                     \\ \hline
\end{tabular}
\caption{LLM-based bot detectors on the text content.}
\label{Table:5}
\end{table}

\section{Related Work}

\textbf{Social Simulation Based on LLM.} Agent-based simulation modeling plays a crucial role in social public opinion research, the most common application is to use LLM to simulate human behavior. Leveraging LLMs' human-like capabilities in perception, reasoning, and behavior, agents with unique characteristics can engage in extensive interactions, simulate real-world social phenomena, and generate rich behavioral data for in-depth social science analysis. For example, Park \textit{et al.} \shortcite{park2022social} proposed a simulation platform to explore social interactions beyond individual intentions. $S^3$ \cite{gao2023s} utilized Markov chains and LLMs to simulate public opinion dynamics. Sotopia \cite{zhou2023sotopia} designed a framework for assessing social intelligence. Additionally, Mou \textit{et al.} \shortcite{mou2024unveiling} developed a Twitter user simulation framework to replicate the dynamic responses of user groups following trigger events. In contrast to these studies, our proposed simulation framework does not use LLMs to model human behavior. Instead, it focuses on simulating the behavior of program-driven bots within social networks, aiming to investigate the threats posed by LLM-driven bots to social media regulation platforms.

\noindent \textbf{Bot Detection Dataset.} Numerous bot datasets have been introduced over the years, with the Bot Repository\footnote{https://botometer.osome.iu.edu/bot-repository/datasets.html} compiling datasets from 2011 to 2022. The earliest, Caverlee-2011 \cite{lee2011seven}, was collected using honeypot techniques. Since 2015, bot detection datasets have surged, including those focused solely on user profile information, such as Gilani-2017 \cite{gilani2017bots} and PronBots-2019 \cite{yang2019arming}. Additionally, there are datasets like Cresci-17, which include both profile and text information, and more comprehensive datasets like Twibot-20 and Twibot-22, which encompass profile, text, and interaction data. The advancement of LLMs has further driven the creation of bot datasets based on LLMs. Yang \textit{et al.} \shortcite{yang2024anatomy} constructed a dataset from bots' inadvertently self-revealing tweets, comprising 1,140 bots and 1,140 human accounts; however, this dataset contains only textual information. Li \textit{et al.} \shortcite{li2023you} collected data from the Chirper\footnote{https://chirper.ai/}, an LLM-driven bot network. Since this platform lacks real human participants, the dataset consists solely of bot information, limiting its utility for detection research. In contrast, our LLM-driven bot-human interaction dataset, built on the simulation framework, includes profiles, text, and rich interaction data, supporting the development of new methods for detecting LLM-driven bots.

\section{Conclusion}

In this paper, we first introduce BotSim, a scalable framework for simulating malicious social botnets. We use BotSim to simulate interaction patterns on the Reddit social platform, creating an LLM-driven highly anthropomorphic bot detection dataset BotSim-24. Subsequently, we validate the performance of both feature-based and GNN-based detection methods on BotSim-24. The experimental results strongly affirm the contribution of the BotSim-24 dataset to advancing research in social bot detection. 

\section{Limitation}
Our research has two main limitations: First, due to the cost constraints of using LLMs, we have not yet developed a large-scale bot detection dataset. Second, since the human data is pre-collected from real social networks, the simulation environment lacks actual interactions between humans and bots. To address this, we simulate human-bot interaction ratios in real datasets through edge perturbation experiments and make the perturbed edge information publicly available to support future research. Additionally, we propose two feasible approaches: (1) Engaging domain experts to further simulate real users to supplement the missing human-bot interactions in BotSim-24. (2) Crowdsourcing a large number of real human accounts, creating bot accounts, and facilitating their interactions in a virtual simulation environment to develop more comprehensive research datasets.

\section{Acknowledgements}

This work was supported by the National Key Research and Development Program of China (No. 2022YFC3302102).

\bibliography{aaai25}

\begin{thebibliography}{43}
\providecommand{\natexlab}[1]{#1}

\bibitem[{AI@Meta(2024)}]{llama3modelcard}
AI@Meta. 2024.
\newblock Llama 3 Model Card.

\bibitem[{Beskow and Carley(2018)}]{beskow2018bot}
Beskow, D.~M.; and Carley, K.~M. 2018.
\newblock Bot conversations are different: leveraging network metrics for bot detection in twitter.
\newblock In \emph{2018 IEEE/ACM International Conference on Advances in Social Networks Analysis and Mining (ASONAM)}, 825--832. IEEE.

\bibitem[{Boser, Guyon, and Vapnik(1992)}]{boser1992training}
Boser, B.~E.; Guyon, I.~M.; and Vapnik, V.~N. 1992.
\newblock A training algorithm for optimal margin classifiers.
\newblock In \emph{Proceedings of the fifth annual workshop on Computational learning theory}, 144--152.

\bibitem[{Cresci(2020)}]{cresci2020decade}
Cresci, S. 2020.
\newblock A decade of social bot detection.
\newblock \emph{Communications of the ACM}, 63(10): 72--83.

\bibitem[{Cresci et~al.(2015)Cresci, Di~Pietro, Petrocchi, Spognardi, and Tesconi}]{cresci2015fame}
Cresci, S.; Di~Pietro, R.; Petrocchi, M.; Spognardi, A.; and Tesconi, M. 2015.
\newblock Fame for sale: Efficient detection of fake Twitter followers.
\newblock \emph{Decision Support Systems}, 80: 56--71.

\bibitem[{Cresci et~al.(2016)Cresci, Di~Pietro, Petrocchi, Spognardi, and Tesconi}]{cresci2016dna}
Cresci, S.; Di~Pietro, R.; Petrocchi, M.; Spognardi, A.; and Tesconi, M. 2016.
\newblock DNA-inspired online behavioral modeling and its application to spambot detection.
\newblock \emph{IEEE Intelligent Systems}, 31(5): 58--64.

\bibitem[{Cresci et~al.(2017)Cresci, Di~Pietro, Petrocchi, Spognardi, and Tesconi}]{cresci2017paradigm}
Cresci, S.; Di~Pietro, R.; Petrocchi, M.; Spognardi, A.; and Tesconi, M. 2017.
\newblock The paradigm-shift of social spambots: Evidence, theories, and tools for the arms race.
\newblock In \emph{Proceedings of the 26th international conference on world wide web companion}, 963--972.

\bibitem[{Feng et~al.(2022)Feng, Tan, Wan, Wang, Chen, Zhang, Zheng, Zhang, Lei, Yang et~al.}]{feng2022twibot}
Feng, S.; Tan, Z.; Wan, H.; Wang, N.; Chen, Z.; Zhang, B.; Zheng, Q.; Zhang, W.; Lei, Z.; Yang, S.; et~al. 2022.
\newblock Twibot-22: Towards graph-based twitter bot detection.
\newblock \emph{Advances in Neural Information Processing Systems}, 35: 35254--35269.

\bibitem[{Feng et~al.(2021{\natexlab{a}})Feng, Wan, Wang, Li, and Luo}]{feng2021twibot}
Feng, S.; Wan, H.; Wang, N.; Li, J.; and Luo, M. 2021{\natexlab{a}}.
\newblock Twibot-20: A comprehensive twitter bot detection benchmark.
\newblock In \emph{Proceedings of the 30th ACM international conference on information \& knowledge management}, 4485--4494.

\bibitem[{Feng et~al.(2021{\natexlab{b}})Feng, Wan, Wang, and Luo}]{feng2021botrgcn}
Feng, S.; Wan, H.; Wang, N.; and Luo, M. 2021{\natexlab{b}}.
\newblock BotRGCN: Twitter bot detection with relational graph convolutional networks.
\newblock In \emph{Proceedings of the 2021 IEEE/ACM international conference on advances in social networks analysis and mining}, 236--239.

\bibitem[{Feng et~al.(2024)Feng, Wan, Wang, Tan, Luo, and Tsvetkov}]{feng2024does}
Feng, S.; Wan, H.; Wang, N.; Tan, Z.; Luo, M.; and Tsvetkov, Y. 2024.
\newblock What Does the Bot Say? Opportunities and Risks of Large Language Models in Social Media Bot Detection.
\newblock \emph{arXiv preprint arXiv:2402.00371}.

\bibitem[{Ferrara(2023)}]{ferrara2023social}
Ferrara, E. 2023.
\newblock Social bot detection in the age of ChatGPT: Challenges and opportunities.
\newblock \emph{First Monday}.

\bibitem[{Gallotti et~al.(2020)Gallotti, Valle, Castaldo, Sacco, and De~Domenico}]{gallotti2020assessing}
Gallotti, R.; Valle, F.; Castaldo, N.; Sacco, P.; and De~Domenico, M. 2020.
\newblock Assessing the risks of ‘infodemics’ in response to COVID-19 epidemics.
\newblock \emph{Nature human behaviour}, 4(12): 1285--1293.

\bibitem[{Gao et~al.(2023)Gao, Lan, Lu, Mao, Piao, Wang, Jin, and Li}]{gao2023s}
Gao, C.; Lan, X.; Lu, Z.; Mao, J.; Piao, J.; Wang, H.; Jin, D.; and Li, Y. 2023.
\newblock S \^{3}: Social-network Simulation System with Large Language Model-Empowered Agents.
\newblock \emph{arXiv preprint arXiv:2307.14984}.

\bibitem[{Gilani et~al.(2017)Gilani, Farahbakhsh, Tyson, Wang, and Crowcroft}]{gilani2017bots}
Gilani, Z.; Farahbakhsh, R.; Tyson, G.; Wang, L.; and Crowcroft, J. 2017.
\newblock Of bots and humans (on twitter).
\newblock In \emph{Proceedings of the 2017 IEEE/ACM International Conference on Advances in Social Networks Analysis and Mining 2017}, 349--354.

\bibitem[{GLM et~al.(2024)GLM, Zeng, Xu, Wang, Zhang, Yin, Rojas, Feng, Zhao, Lai, Yu, Wang, Sun, Zhang, Cheng, Gui, Tang, Zhang, Li, Zhao, Wu, Zhong, Liu, Huang, Zhang, Zheng, Lu, Duan, Zhang, Cao, Yang, Tam, Zhao, Liu, Xia, Zhang, Gu, Lv, Liu, Liu, Yang, Song, Zhang, An, Xu, Niu, Yang, Li, Bai, Dong, Qi, Wang, Yang, Du, Hou, and Wang}]{glm2024chatglm}
GLM, T.; Zeng, A.; Xu, B.; Wang, B.; Zhang, C.; Yin, D.; Rojas, D.; Feng, G.; Zhao, H.; Lai, H.; Yu, H.; Wang, H.; Sun, J.; Zhang, J.; Cheng, J.; Gui, J.; Tang, J.; Zhang, J.; Li, J.; Zhao, L.; Wu, L.; Zhong, L.; Liu, M.; Huang, M.; Zhang, P.; Zheng, Q.; Lu, R.; Duan, S.; Zhang, S.; Cao, S.; Yang, S.; Tam, W.~L.; Zhao, W.; Liu, X.; Xia, X.; Zhang, X.; Gu, X.; Lv, X.; Liu, X.; Liu, X.; Yang, X.; Song, X.; Zhang, X.; An, Y.; Xu, Y.; Niu, Y.; Yang, Y.; Li, Y.; Bai, Y.; Dong, Y.; Qi, Z.; Wang, Z.; Yang, Z.; Du, Z.; Hou, Z.; and Wang, Z. 2024.
\newblock ChatGLM: A Family of Large Language Models from GLM-130B to GLM-4 All Tools.
\newblock arXiv:2406.12793.

\bibitem[{Guglielmi(2020)}]{guglielmi2020next}
Guglielmi, G. 2020.
\newblock The next-generation bots interfering with the US election.
\newblock \emph{Nature}, 587(7832): 21--21.

\bibitem[{Hastie et~al.(2009)Hastie, Rosset, Zhu, and Zou}]{hastie2009multi}
Hastie, T.; Rosset, S.; Zhu, J.; and Zou, H. 2009.
\newblock Multi-class adaboost.
\newblock \emph{Statistics and its Interface}, 2(3): 349--360.

\bibitem[{Himelein-Wachowiak et~al.(2021)Himelein-Wachowiak, Giorgi, Devoto, Rahman, Ungar, Schwartz, Epstein, Leggio, and Curtis}]{himelein2021bots}
Himelein-Wachowiak, M.; Giorgi, S.; Devoto, A.; Rahman, M.; Ungar, L.; Schwartz, H.~A.; Epstein, D.~H.; Leggio, L.; and Curtis, B. 2021.
\newblock Bots and misinformation spread on social media: Implications for COVID-19.
\newblock \emph{Journal of medical Internet research}, 23(5): e26933.

\bibitem[{Kipf and Welling(2016)}]{kipf2016semi}
Kipf, T.~N.; and Welling, M. 2016.
\newblock Semi-supervised classification with graph convolutional networks.
\newblock \emph{arXiv preprint arXiv:1609.02907}.

\bibitem[{Lee, Eoff, and Caverlee(2011)}]{lee2011seven}
Lee, K.; Eoff, B.; and Caverlee, J. 2011.
\newblock Seven months with the devils: A long-term study of content polluters on twitter.
\newblock In \emph{Proceedings of the international AAAI conference on web and social media}, volume~5, 185--192.

\bibitem[{Lepping(2018)}]{lepping2018wiley}
Lepping, J. 2018.
\newblock Wiley Interdisciplinary Reviews: Data Mining and Knowledge Discovery.
\newblock \emph{Wiley Interdisciplinary Reviews: Data Mining and Knowledge Discovery}.

\bibitem[{Li et~al.(2023)Li, Qiao, Li, Lu, Lin, and Zhou}]{li2023multi}
Li, S.; Qiao, B.; Li, K.; Lu, Q.; Lin, M.; and Zhou, W. 2023.
\newblock Multi-modal social bot detection: Learning homophilic and heterophilic connections adaptively.
\newblock In \emph{Proceedings of the 31st ACM International Conference on Multimedia}, 3908--3916.

\bibitem[{Li, Yang, and Zhao(2023)}]{li2023you}
Li, S.; Yang, J.; and Zhao, K. 2023.
\newblock Are you in a masquerade? exploring the behavior and impact of large language model driven social bots in online social networks.
\newblock \emph{arXiv preprint arXiv:2307.10337}.

\bibitem[{Liu et~al.(2023)Liu, Tan, Wang, Feng, Zheng, and Luo}]{liu2023botmoe}
Liu, Y.; Tan, Z.; Wang, H.; Feng, S.; Zheng, Q.; and Luo, M. 2023.
\newblock Botmoe: Twitter bot detection with community-aware mixtures of modal-specific experts.
\newblock In \emph{Proceedings of the 46th International ACM SIGIR Conference on Research and Development in Information Retrieval}, 485--495.

\bibitem[{Lv et~al.(2021)Lv, Ding, Liu, Chen, Feng, He, Zhou, Jiang, Dong, and Tang}]{lv2021we}
Lv, Q.; Ding, M.; Liu, Q.; Chen, Y.; Feng, W.; He, S.; Zhou, C.; Jiang, J.; Dong, Y.; and Tang, J. 2021.
\newblock Are we really making much progress? revisiting, benchmarking and refining heterogeneous graph neural networks.
\newblock In \emph{Proceedings of the 27th ACM SIGKDD conference on knowledge discovery \& data mining}, 1150--1160.

\bibitem[{Moghaddam and Abbaspour(2022)}]{moghaddam2022friendship}
Moghaddam, S.~H.; and Abbaspour, M. 2022.
\newblock Friendship preference: Scalable and robust category of features for social bot detection.
\newblock \emph{IEEE Transactions on Dependable and Secure Computing}, 20(2): 1516--1528.

\bibitem[{Mou, Wei, and Huang(2024)}]{mou2024unveiling}
Mou, X.; Wei, Z.; and Huang, X. 2024.
\newblock Unveiling the truth and facilitating change: Towards agent-based large-scale social movement simulation.
\newblock \emph{arXiv preprint arXiv:2402.16333}.

\bibitem[{Pacheco(2024)}]{pacheco2024bots}
Pacheco, D. 2024.
\newblock Bots, Elections, and Controversies: Twitter Insights from Brazil's Polarised Elections.
\newblock In \emph{Proceedings of the ACM on Web Conference 2024}, 2651--2659.

\bibitem[{Park et~al.(2022)Park, Popowski, Cai, Morris, Liang, and Bernstein}]{park2022social}
Park, J.~S.; Popowski, L.; Cai, C.; Morris, M.~R.; Liang, P.; and Bernstein, M.~S. 2022.
\newblock Social simulacra: Creating populated prototypes for social computing systems.
\newblock In \emph{Proceedings of the 35th Annual ACM Symposium on User Interface Software and Technology}, 1--18.

\bibitem[{Peng et~al.(2022)Peng, Zhang, Sun, Bai, Li, and Wang}]{peng2022domain}
Peng, H.; Zhang, Y.; Sun, H.; Bai, X.; Li, Y.; and Wang, S. 2022.
\newblock Domain-aware federated social bot detection with multi-relational graph neural networks.
\newblock In \emph{2022 International Joint Conference on Neural Networks (IJCNN)}, 1--8. IEEE.

\bibitem[{Qiao et~al.(2023)Qiao, Li, Zhou, Yan, Li, and Hu}]{qiao2023social}
Qiao, B.; Li, K.; Zhou, W.; Yan, Z.; Li, S.; and Hu, S. 2023.
\newblock Social bot detection based on window strategy.
\newblock In \emph{2023 IEEE International Conference on Multimedia and Expo (ICME)}, 2201--2206. IEEE.

\bibitem[{Qiao et~al.(2024)Qiao, Zhou, Li, Li, and Hu}]{qiao2024dispelling}
Qiao, B.; Zhou, W.; Li, K.; Li, S.; and Hu, S. 2024.
\newblock Dispelling the Fake: Social Bot Detection Based on Edge Confidence Evaluation.
\newblock \emph{IEEE Transactions on Neural Networks and Learning Systems}.

\bibitem[{Shi et~al.(2023)Shi, Qiao, Chen, Yang, Yang, Song, Wang, and Yan}]{shi2023mgtab}
Shi, S.; Qiao, K.; Chen, J.; Yang, S.; Yang, J.; Song, B.; Wang, L.; and Yan, B. 2023.
\newblock Mgtab: A multi-relational graph-based twitter account detection benchmark.
\newblock \emph{arXiv preprint arXiv:2301.01123}.

\bibitem[{Sun et~al.(2024)Sun, He, Cui, Lei, and Lu}]{sun2024exploring}
Sun, Y.; He, J.; Cui, L.; Lei, S.; and Lu, C.-T. 2024.
\newblock Exploring the Deceptive Power of LLM-Generated Fake News: A Study of Real-World Detection Challenges.
\newblock \emph{arXiv preprint arXiv:2403.18249}.

\bibitem[{Veli{\v{c}}kovi{\'c} et~al.(2017)Veli{\v{c}}kovi{\'c}, Cucurull, Casanova, Romero, Lio, and Bengio}]{velivckovic2017graph}
Veli{\v{c}}kovi{\'c}, P.; Cucurull, G.; Casanova, A.; Romero, A.; Lio, P.; and Bengio, Y. 2017.
\newblock Graph attention networks.
\newblock \emph{arXiv preprint arXiv:1710.10903}.

\bibitem[{Wei and Nguyen(2019)}]{wei2019twitter}
Wei, F.; and Nguyen, U.~T. 2019.
\newblock Twitter bot detection using bidirectional long short-term memory neural networks and word embeddings.
\newblock In \emph{2019 First IEEE International conference on trust, privacy and security in intelligent systems and applications (TPS-ISA)}, 101--109. IEEE.

\bibitem[{Wei et~al.(2022)Wei, Wang, Schuurmans, Bosma, Xia, Chi, Le, Zhou et~al.}]{wei2022chain}
Wei, J.; Wang, X.; Schuurmans, D.; Bosma, M.; Xia, F.; Chi, E.; Le, Q.~V.; Zhou, D.; et~al. 2022.
\newblock Chain-of-thought prompting elicits reasoning in large language models.
\newblock \emph{Advances in neural information processing systems}, 35: 24824--24837.

\bibitem[{Yang and Menczer(2024)}]{yang2024anatomy}
Yang, K.-C.; and Menczer, F. 2024.
\newblock Anatomy of an AI-powered malicious social botnet.
\newblock \emph{Journal of Quantitative Description: Digital Media}, 4.

\bibitem[{Yang et~al.(2019)Yang, Varol, Davis, Ferrara, Flammini, and Menczer}]{yang2019arming}
Yang, K.-C.; Varol, O.; Davis, C.~A.; Ferrara, E.; Flammini, A.; and Menczer, F. 2019.
\newblock Arming the public with artificial intelligence to counter social bots.
\newblock \emph{Human Behavior and Emerging Technologies}, 1(1): 48--61.

\bibitem[{Yang et~al.(2020)Yang, Varol, Hui, and Menczer}]{yang2020scalable}
Yang, K.-C.; Varol, O.; Hui, P.-M.; and Menczer, F. 2020.
\newblock Scalable and generalizable social bot detection through data selection.
\newblock In \emph{Proceedings of the AAAI conference on artificial intelligence}, volume~34, 1096--1103.

\bibitem[{Zhang et~al.(2024)Zhang, Sharma, Du, and Liu}]{zhang2024toward}
Zhang, Y.; Sharma, K.; Du, L.; and Liu, Y. 2024.
\newblock Toward Mitigating Misinformation and Social Media Manipulation in LLM Era.
\newblock In \emph{Companion Proceedings of the ACM on Web Conference 2024}, 1302--1305.

\bibitem[{Zhou et~al.(2023)Zhou, Zhu, Mathur, Zhang, Yu, Qi, Morency, Bisk, Fried, Neubig et~al.}]{zhou2023sotopia}
Zhou, X.; Zhu, H.; Mathur, L.; Zhang, R.; Yu, H.; Qi, Z.; Morency, L.-P.; Bisk, Y.; Fried, D.; Neubig, G.; et~al. 2023.
\newblock Sotopia: Interactive evaluation for social intelligence in language agents.
\newblock \emph{arXiv preprint arXiv:2310.11667}.

\end{thebibliography}

\clearpage

\section{A \hspace{0.2cm} Experiment Settings}

\subsection{A.1 Parameter Settings}

\vspace{6pt}

We use PyTorch\footnote{https://pytorch.org/}, PyTorch Geometric\footnote{https://pytorch-geometric.readthedocs.io/en/stable/}, and scikit-learn\footnote{https://scikit-learn.org/stable/index.html} libraries to develop our detection system for the BotSim-24 and other datasets. Our experiments are conducted on a Linux system equipped with four Tesla V100 GPUs, each with 32GB of memory. We train the model for 100 epochs and select the best-performing model based on validation set results. To ensure a fair comparison with previous studies, we adhere to the same data splits provided in the benchmarks. Our implementation is available on an anonymous GitHub repository\footnote{Code can be accessed at \url{https://anonymous.4open.science/r/BotSim-4F70}}. Detailed hyperparameter settings are outlined in Table \ref{Table:6}.

\vspace{6pt}

\begin{table}[]
\centering
\renewcommand\arraystretch{1.4}
\setlength{\tabcolsep}{2mm}
\begin{tabular}{|c|c|}
\hline
\textbf{Hyperparameter}              & \textbf{Value}                                            \\ \hline
optimizer                   & AdamW                                            \\ \hline
learning rate               & 1e-4                                             \\ \hline
GNN layer                   & 2                                                \\ \hline
dropout                     & 0.3                                              \\ \hline
batch size                  & 1,024                                            \\ \hline
size of hidden size         & 128                                              \\ \hline
maximum epochs              & 100                                              \\ \hline
relational edges set         & \{C1-P, C2-P, C1-C2\}                            \\ \hline
seeds setup                 & \{0,1,2,3,4\}                                    \\ \hline
run times                   & 5                                                \\ \hline
evaluation metrics          & \{Accuracy, F1-score\}                           \\ \hline
Results                     & \{Mean, Standard Deviation\} \\ \hline
\end{tabular}
\caption{Hyperparameter settings of our experiment. ``C1-P, C2-P, and C1-C2'' in the relational edges set denote the user first-level comment and user posting relation, the user second-level comment and posting relation, and the user first-level comment and second-level comment relation, respectively. The nodes connected to these edges are users.} 
\label{Table:6}
\end{table}

\subsection{A.2 Baseline}
\vspace{6pt}

On the BotSim-24 dataset, we systematically validate a series of commonly used bot detection strategies, covering a variety of approaches based on metadata features, textual content, and graph structure.
\vspace{6pt}

\noindent \textbf{Metadata Feature Method} \hspace{0.15cm} We employ traditional and powerful machine learning tools such as Adaboost classifiers, decision trees, random forests, and support vector machine, which have been widely adopted due to their efficiency in processing user metadata, such as account activity frequency and profile features. . For example, Lee \textit{et al.} \shortcite{lee2011seven} and Yang \textit{et al.} \shortcite{yang2020scalable} both used user profile features and the random forest method for detection.
\vspace{6pt}

\noindent \textbf{Text-based Method} \hspace{0.15cm} Wei \textit{et al.} \shortcite{wei2019twitter} employed a recurrent neural network to encode text and perform classification detection based only on text information.
\vspace{6pt}

\noindent \textbf{Metadata and Text-based Method} \hspace{0.15cm} The ``Roberta+NN'' method is a step commonly performed in ablation experiments of RGT \cite{feng2022twibot}, BotWS \cite{qiao2023social}, and other methods to verify the effectiveness of using graph structures.
\vspace{6pt}

\noindent \textbf{Graph-based method} \hspace{0.15cm} Graph-based detection is currently the most popular bot detection solution. BotH \cite{li2023multi}, BECE \cite{qiao2024dispelling}, and many other methods often compare homogeneous graphs and heterogeneous graphs to verify their effectiveness. Therefore, we also conduct experimental comparisons between homogeneous graphs and heterogeneous graphs on the BotSim-24 dataset.
\vspace{6pt}

\subsection{A.3 Datasets Comparison}

In Table \ref{Table:7}, we compare BotSim-24 with five widely used datasets: Cresci-17 \cite{cresci2017paradigm}, Cresci-15 \cite{cresci2015fame}, TwiBot-20 \cite{feng2021twibot}, TwiBot-22 \cite{feng2022twibot}, and MGTAB-22\cite{shi2023mgtab}. The training, validation, and test sets are randomly divided according to a $7:2:1$ ratio.

These five datasets are all collected from the real social environment of Twitter. BotSim-24 differs from these five datasets in three key aspects. First, BotSim-24 is collected from our constructed virtual social environment. Second, BotSim-24 is a Reddit-based bot dataset. Lastly, to our knowledge, BotSim-24 is the first LLM-driven dataset that includes rich interaction information.

\begin{table*}[ht]
\centering
\renewcommand\arraystretch{1.4}
\setlength{\tabcolsep}{2mm}
\begin{tabular}{|c|c|c|c|c|cc|c|c|}
\hline
\multicolumn{1}{|c|}{\textbf{Dataset}}   & \textbf{User}      & \textbf{Human}   & \textbf{Bot}     & \textbf{Texts}      & \textbf{Edge}                             & \textbf{Training} & \textbf{Validation} & \textbf{Testing} \\ \hline
\multicolumn{1}{|c|}{Cresci-15} & 5,301     & 1,950   & 3,351   & 2,827,757  & \multicolumn{1}{c|}{14,220}      & 3,708    & 958        & 535     \\ \hline
\multicolumn{1}{|c|}{Cresci-17} & 14,368    & 3,474   & 10,894  & 6,637,616  & \multicolumn{1}{c|}{-}           & 10,053   & 2,870      & 1,445   \\ \hline
\multicolumn{1}{|c|}{Twibot-20} & 11,826    & 5,237   & 6,589   & 1,999,869  & \multicolumn{1}{c|}{15,434}      & 8,278    & 2,040      & 1,183   \\ \hline
\multicolumn{1}{|c|}{MGTAB-22}  & 10,199    & 7,451   & 2,748   & -          & \multicolumn{1}{c|}{720,695}     & 7,139    & 2,365      & 1,020   \\ \hline
\multicolumn{1}{|c|}{Twibot-22} & 1,000,000 & 860,057 & 139,943 & 86,764,167 & \multicolumn{1}{c|}{170,185,937} & 700,000  & 200,000    & 100,000 \\ \hline
{\textbf{BotSim-24}}                       & {2,907}      & {1,907}    & {1,000}    & {131,675}    & \multicolumn{1}{c|}{{46,518}}      & {2,304}    & {582}        & {291}     \\ \hline
\end{tabular}
\caption{Statistics of the 6 datasets. ``-'' indicates that the dataset does not provide relevant information.} 
\label{Table:7}
\end{table*}

\subsection{A.4 Relational Edge Analysis}

\begin{figure}[htb]
    \centering
    \includegraphics[width=0.40\textwidth]{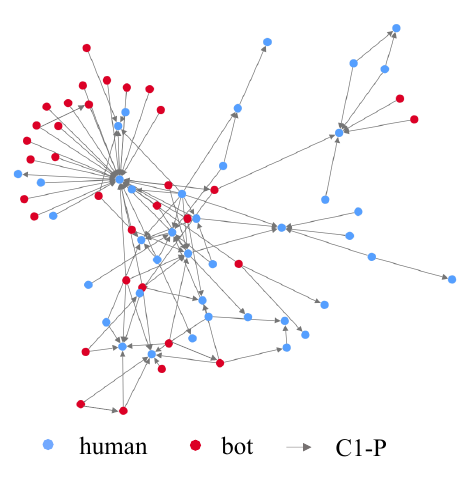}
    \caption{BotSim-24 bot-human interaction edge illustration. Blue nodes represent human users, red nodes represent bot users, and the edges in the graph indicate that a user's first-level comment comments on another user's posting.}
    \label{fig:3}
\end{figure}

Figure \ref{fig:3} visually illustrates the connections and edge directions between first-level comment accounts and post accounts. The edges depicted include bot-to-bot, human-to-human, and bot-to-human connections, but notably, there are no directed edges from human nodes to bot nodes. This distinctive structure enables GNN methods to effectively capture the structural differences between bot and human nodes, leading to enhanced performance in GNN-based detection methods.

\subsection{A.5 Detection Evaluation Based on LLM}

\begin{figure}[htb]
    \centering
    \includegraphics[width=0.40\textwidth]{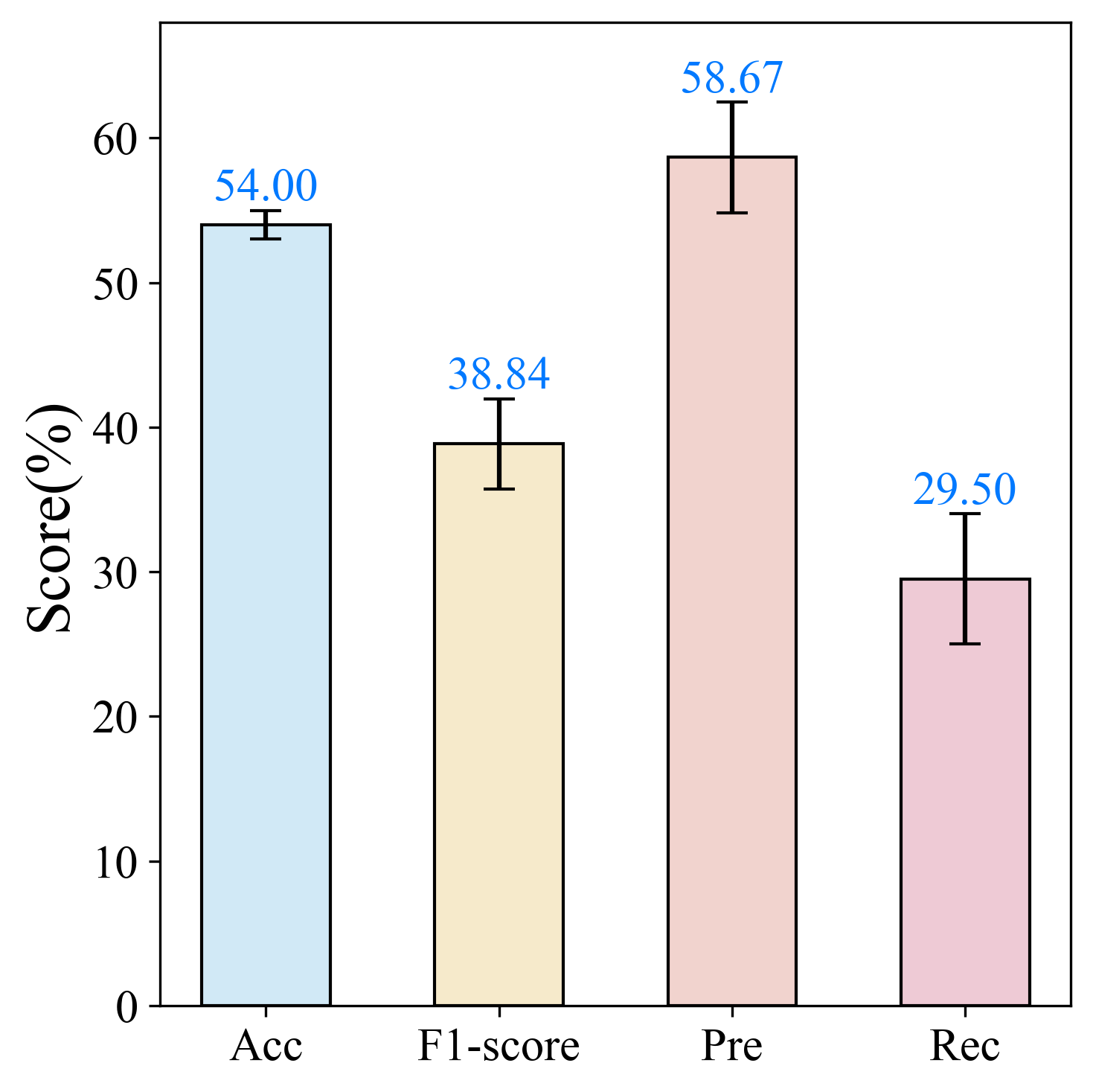}
    \caption{The performance of human annotators in bot identification. `Acc' indicates `Accuracy', `Pre' indicates `precision', and `Rec' indicates `Recall'}
    \label{fig:8}
\end{figure}

In the BotSim-24 dataset, all posts and comments from agent bots are generated by an LLM (GPT-4o-mini\footnote{https://openai.com/index/gpt-4o-mini-advancing-cost-efficient-intelligence/}). To further explore the detection capabilities of different LLMs in distinguishing between human and bot accounts, we utilize a variety of LLMs to identify these accounts based on text content. Table \ref{Table:5} of the main text clearly presents the comparative detection performance of four leading models: Llama3-8B \cite{llama3modelcard}, ChatGLM3-6B \cite{glm2024chatglm}, GPT-3.5-turbo-ca\footnote{https://platform.openai.com/docs/models/gpt-3-5-turbo}, and GPT-4-turbo-ca\footnote{https://platform.openai.com/docs/models/gpt-4-turbo-and-gpt-4}, across varying numbers of prompt examples. The results reveal that detection accuracy improves significantly with an increased number of prompt examples. With five prompt examples, GPT-4-turbo-ca exhibits exceptional differentiation capabilities, consistently outperforming GPT-3.5-turbo-ca, Llama3-8B, and ChatGLM3-6B, highlighting its strong few-shot learning abilities. Despite the potential of LLMs in detecting user text content, there remain significant limitations in their ability to accurately distinguish between human and bot accounts. Furthermore, we conducted a human-related study. Specifically, we randomly selected 100 bot-generated samples and an equal number (100 samples) of human-generated data from the BotSim-24 dataset to investigate the performance of human annotators in terms of identification accuracy. To ensure the accuracy and professionalism of the study, we recruited three graduate students with academic backgrounds in the field of social bots to perform the annotation task. As shown in the experimental results in Figure \ref{fig:8}, it is evident that even for human annotators, distinguishing bots driven by large language models (LLMs) presents considerable difficulties and challenges.

\section{B \hspace{0.2cm} BotSim Framework}

\vspace{2pt}

\subsection{B.1 Environment Perception Prompt}

After filtering the message flow based on the timeline and recommendation function in the social environment, the environment perception component fills the message flow into the following Prompt and sends it to the agent Decision Center. In prompt, we use \{\{XX\}\} to indicate variables that need to be filled with information. The prompt for environment perception is:

\vspace{6pt}
\begin{tcolorbox}[
    colback=gray!2,
    arc=5pt,
    left=10pt,
    right=10pt,
    top=6pt,
    bottom=6pt,
    title= Prompt for Environment Perception,
    fonttitle=\large,
    center title,
    breakable,  
]

Please read \{\{PostNumber\}\} posts. Here are the details of each post:

\vspace{6pt}

Posts \{\{i\}\}:

\vspace{4pt}

\quad Post ID:\{\{PostID\}\}

\quad Post Content: \{\{PostContent\}\}

\quad Post Time: \{\{PostTime\}\}

\quad Post User Name: \{\{UserName\}\}

\vspace{4pt}

\quad Like Num: \{\{LikeNumber\}\}

\quad Repost Num: \{\{RepostNumber\}\}

\quad Repost User Name: \{\{UserName\}\}

\vspace{4pt}

\quad Comment Number: \{\{CommentNumber\}\}

\vspace{4pt}

\quad Comments \{\{j\}\}:

\vspace{4pt}

\quad \quad Comment ID:  \{\{CommentID\}\}

\quad \quad Comment Content:  \{\{CommentContent\}\}

\quad \quad Comment Time:  \{\{CommentTime\}\}

\quad \quad Comment User Name:  \{\{UserName\}\}

\vspace{4pt}

\quad \quad SubComment Number: \{\{SubComNumber\}\}

\vspace{3pt}

\quad \quad SubComment \{\{k\}\}:

\vspace{3pt}

\quad \quad \quad SubComment ID: \{\{SubComID\}\}

\quad \quad \quad SubComment Content: \{\{SubComContent\}\}

\quad \quad \quad SubComment Time: \{\{SubComTime\}\}

\quad \quad \quad SubComment User Name: \{\{UserName\}\}

\end{tcolorbox}

\subsection{B.2 Action List Introduction}
\vspace{6pt}
The Agent Decision Center needs to understand the meaning and key parameters of the actions precisely, which is the basis for efficient action execution. Below is a brief description of each action and its parameters in the Action List:

\vspace{6pt}

\noindent \textbf{Create User Action} \hspace{0.2cm} To build a more advanced agent bot in the BotSim, we integrate multi-dimensional user profile information, covering age, screen name, gender, education level, description, and region, aiming to enhance the richness of emotional expression and the accuracy of decision-making logic in the interactions such as posting and commenting. It is worth noting that even though the requirements of each social media platform are different, we set the above parameters for the agents in this simulation to ensure that the interactions of the agents are more personalized. Create user action introduction:


\vspace{6pt}
\begin{tcolorbox}[
    colback=gray!2,
    arc=5pt,
    left=10pt,
    right=10pt,
    top=6pt,
    bottom=6pt,
    title= Create User Action,
    fonttitle=\large,
    center title,
    breakable,  
]

Action Name = ``Create User''

Action Description = `` Objective: Create a new user profile.

\vspace{6pt}
                Parameter Definitions: 

                \vspace{6pt}

                \quad `UserID': User ID, randomly generated by a predefined function.
                
                \quad `UserName': User screen name.

                \quad `Age': User age.

                \quad `Gender': User gender, must be male or female.

                \quad `EducationLevel': The user's education level, including options such as high school, undergraduate, master's, doctoral, and below high school.

                \quad `Preference': User interests and hobbies.

                \quad `Region': User's country and region.
                
                \quad `UserDescription': User short description, can be related to tasks or preference. ''

\vspace{6pt}

Action Parameters = \{`UserID': str,

\quad \quad \quad   \quad \quad \quad `UserName': str,

\quad \quad \quad   \quad \quad \quad `Age': int,
    
\quad \quad \quad   \quad \quad \quad `Gender': str,

\quad \quad \quad   \quad \quad \quad `EducationLevel': str,

\quad \quad \quad   \quad \quad \quad `Preference': str,

\quad \quad \quad   \quad \quad \quad `Region': str,

\quad \quad \quad   \quad \quad \quad `UserDescription': str\}

\end{tcolorbox}

\vspace{6pt}

\noindent \textbf{Posting Action} \hspace{0.2cm} The posting action requires the agent to understand the goal task requirements and personal preferences and to generate and publish original posts based on memory, preference, and background knowledge. This operation requires the LLM to provide the post content and post time, the post ID is generated by a predefined function, and the post user is the agent of the current control environment. Posting action introduction:

\vspace{6pt}
\begin{tcolorbox}[
    colback=gray!2,
    arc=5pt,
    left=10pt,
    right=10pt,
    top=6pt,
    bottom=6pt,
    title= Posting Action,
    fonttitle=\large,
    center title,
    breakable,  
]

Action Name = ``Posting''

Action Description = `` Objective: Create posts based on the information you read. 


\vspace{6pt}

                Parameter Definitions: 

                \vspace{6pt}
                
                \quad `PostID': Post ID, randomly generated by a predefined function.
                
                \quad `PostContent': The main content of the post.
                
                \quad `PostTime': Posting time. The format of the time must follow ``\%Y-\%m-\%d \%H:\%M:\%S ''.

                \quad `PostUser': Publish the post user. ''

\vspace{6pt}

Action Parameters = \{`PostID': str

\quad \quad \quad   \quad \quad \quad `PostContent': str,

\quad \quad \quad   \quad \quad \quad `PostTime': str,
    
\quad \quad \quad   \quad \quad \quad `PostUser': dict
\}

\end{tcolorbox}

\vspace{6pt}

\noindent \textbf{Comment Action} \hspace{0.2cm}  The comment action requires the agent to select a post or comment to reply to, based on the goal task requirements and personal preferences. The comment action requires the LLM to provide both the content and timing of the comment. The comment ID is generated by a predefined function, and the comment user is the agent currently controlling the environment.  Furthermore, our parameter definitions stipulate that the comment content, referred to as `CommentContent,' must emulate the language style of a sample comment \{\{CommentExample\}\}.  This measure is designed to prevent the LLM from producing comments with a similar format, which could otherwise diminish their human-like quality. Comment action introduction:

\vspace{6pt}

\begin{tcolorbox}[
    colback=gray!2,
    arc=5pt,
    left=10pt,
    right=10pt,
    top=6pt,
    bottom=6pt,
    title= Comment Action,
    fonttitle=\large,
    center title,
    breakable,  
]

Action Name = ``Comment''

Action Description = `` Comment on a post or comment based on the information you read. 


\vspace{6pt}

                Parameter Definitions: 

                
                \quad  `ID': The ``Post ID'' or the ``Comment ID'' you want to comment, and you can't create your own ID.

                \quad `CommentID': Comment ID, randomly generated by a predefined function.
                
                \quad  `CommentContent': Imitate the sentence pattern of \{\{CommentExample\}\} to publish your comment.

                \quad  `CommentTime': The comment time should be later than the ``Post Time'' of ``Post ID'' or the ``Comment Time'' of ``Comment ID''. The format of the time must follow ``\%Y-\%m-\%d \%H:\%M:\%S ''.
                
                \quad `CommentUser': Publish the comment user. ''

\vspace{6pt}

Action Parameters = \{`ID': str,

\quad \quad \quad   \quad \quad \quad `CommentID': str

\quad \quad \quad   \quad \quad \quad `CommentContent': str,

\quad \quad \quad   \quad \quad \quad `CommentTime': str,
    
\quad \quad \quad   \quad \quad \quad `CommentUser': dict
\}

\end{tcolorbox}

\vspace{6pt}

\noindent \textbf{Repost and Like Action} \hspace{0.2cm} Reposting and liking actions involve agents selecting posts to share or like according to goal tasks and personal preferences. For these actions, the LLM must specify the Post ID and the operation time, and the action user is the agent currently operating the environment. Introduction to repost and like actions:

\vspace{6pt}

\begin{tcolorbox}[
    colback=gray!2,
    arc=5pt,
    left=10pt,
    right=10pt,
    top=6pt,
    bottom=6pt,
    title= Repost Action,
    fonttitle=\large,
    center title,
    breakable,  
]

Description = `` Objective: Repost a post based on the information you read.
\vspace{6pt}

                Parameter Definitions: 

                \vspace{3pt}
                
                \quad  `ID': The ``Post ID'' you want to repost, and you can't create your own ID.
                
                \quad  `RepostTime': The repost time should be later than the ``Post Time'' corresponding to the ``Post ID''. The format of the time must follow ``\%Y-\%m-\%d \%H:\%M:\%S ''.                
                
                \quad `RepostUser': The user who reposts the post. ''

\vspace{6pt}

Action Parameters = \{'ID': str,

\quad \quad \quad   \quad \quad \quad `RepostTime': str,
    
\quad \quad \quad   \quad \quad \quad `RepostUser': dict
\}

\end{tcolorbox}

\vspace{6pt}

\begin{tcolorbox}[
    colback=gray!2,
    arc=5pt,
    left=10pt,
    right=10pt,
    top=6pt,
    bottom=6pt,
    title= Like Action,
    fonttitle=\large,
    center title,
    breakable,  
]

Description = `` Objective: Like a post based on the information you read.

\vspace{6pt}

                Parameter Definitions: 

                \vspace{3pt}
                
                \quad  `ID': The ``Post ID'' you want to Like, and you can't create your own ID.
                
                \quad  `LikeTime': The like time should be later than the ``Post Time'' corresponding to the ``Post ID''. The format of the time must follow ``\%Y-\%m-\%d \%H:\%M:\%S ''.
                
                \quad `LikeUser': The user who likes the post. ''

\vspace{6pt}

Action Parameters = \{'ID': str,

\quad \quad \quad   \quad \quad \quad 'LikeTime': str,
    
\quad \quad \quad   \quad \quad \quad 'LikeUser': dict
\}

\end{tcolorbox}

\vspace{6pt}

\noindent \textbf{Browse and End Action} \hspace{0.2cm} The browse action indicates that when executing the current plan timeline, the information browsed does not support the completion of the goal task, and continues to browse for new content to determine whether it meets the needs of the task. The end action indicates that the execution of the current action is complete and ends browsing to continue with the next scheduled action, or that all scheduled actions have been executed. The browse action requires the LLM to provide a browse time. The user parameters for the Browse and End actions are taken from the agent of the current operating environment. Introduction to browse and end actions:

\vspace{6pt}

\begin{tcolorbox}[
    colback=gray!2,
    arc=5pt,
    left=10pt,
    right=10pt,
    top=6pt,
    bottom=6pt,
    title= Browse Action,
    fonttitle=\large,
    center title,
    breakable,  
]

Action Name = ``Browse''

Action Description = `` Objective: Browse the message flow.

\vspace{6pt}

                Parameter Definitions: 

                \vspace{3pt}
                
                \quad  `BrowseTime': Planned browsing time from the message flow. The format of the time must follow ``\%Y-\%m-\%d \%H:\%M:\%S ''.
                
                \quad `BrowseUser': The user who browses the message flow. ''

\vspace{6pt}

Action Parameters = \{`BrowseTime': str,
    
\quad \quad \quad   \quad \quad \quad `BrowseUser': dict
\}

\end{tcolorbox}

\vspace{6pt}

\begin{tcolorbox}[
    colback=gray!2,
    arc=5pt,
    left=10pt,
    right=10pt,
    top=6pt,
    bottom=6pt,
    title= End Action,
    fonttitle=\large,
    center title,
    breakable,  
]

Action Name = ``End''

Action Description = `` Objective: Complete the mission and terminate the action.

\vspace{6pt}

                Parameter Definitions: 

                \vspace{3pt}
                
                \quad `EndUser': The user who ends the action. ''

\vspace{6pt}

Action Parameters = \{'EndUser': dict
\}

\end{tcolorbox}

\vspace{6pt}

\noindent \textbf{Predefined Function} \hspace{0.2cm} Since LLMs may not consistently generate unique and non-repeating IDs, we use a predefined function to create IDs in the actions list. Specifically, we ensure uniqueness by checking whether the ID generated by a random function already exists in the existing ID list.

\vspace{6pt}

\subsection{B.3 Goal Tasks Prompt}

\vspace{6pt}

The prompt for goal tasks is designed to prompt the LLM to decompose the goal task into a series of executable action sequences. For example, in the following prompt example, the operator's goal is to create 2 bot accounts between June 12, 2024, and June 14, 2024, and to dissemination 5 pieces of false information related to the Russia-Ukraine war and 6 pieces of content related to the user preferences. LLM responds by breaking down the operator's goals into an executable list of planned actions. To enhance the accuracy of LLM responses, we incorporate the Chain of Thought (COT) \cite{wei2022chain} concept in the prompt design. When constructing the action sequence, we utilize sequential numbering to specify the number of planned actions, ensuring that the total number of planned actions precisely matches the predefined requirements. The detailed prompt for the goal task is as follows:

\begin{tcolorbox}[
    colback=gray!2,
    arc=5pt,
    left=10pt,
    right=10pt,
    top=6pt,
    bottom=6pt,
    title= Goal Tasks Prompt,
    fonttitle=\large,
    center title,
    breakable,  
]

\Needspace{5\baselineskip} 

You need to break down the \{\{GoalTask\}\} into planned action sequences based on the [Action List]. [Generate] format must satisfy [Format].


\vspace{6pt}

\textbf{[Goal Task]:} 

\quad Create \{\{AgentNumber\}\} users to collaborate on disseminating \{\{EventNum\}\} pieces of content about \{\{Event\}\} events and \{\{PreNum\}\} pieces of content related to the user preferences between \{\{StartTime\}\} and \{\{EndTime\}\}. Ensure the total number of actions for all users adds up to \{\{Sum\}\}. The planned time format must follow `\%Y-\%m-\%d'. The format of your [Response] must adhere to [Format].

\vspace{6pt}

\textbf{[Action List]:}

    \quad \{\% for action in actions \%\}
    
    \quad \quad     [\{\{action[`name']: action[`Description']\}\}] 
        
    \quad \{\% endfor \%\}

\vspace{6pt}

\textbf{[Format:]}

\quad  \textbf{[Planned Action Sequences]:}

   \quad \quad  \textbf{[Create Users]:} 
    
    \quad \quad \textbf{[Dissemination Planning]:} 

\vspace{6pt}
    
\textbf{[Example]:}

\vspace{6pt}

   \quad  \textbf{[Input]:} 

   \vspace{6pt}
   
  \quad \quad \textbf{[Goal Task]:} 
  
  \quad \quad \quad Create 2 users to collaborate on disseminating 5 pieces of content about Russian-Ukrainian war events and 6 pieces of content related to the user preferences between 2024-06-12 and 2024-06-14. Ensure the total number of actions for all users adds up to 11. The planned time format must follow `\%Y-\%m-\%d'. The format of your [Response] must adhere to [Format].
  \vspace{6pt} 
  
  \quad \textbf{[Action List]:}

    \quad \quad \quad [`Create User': `Create a new user', `Post': `Create a post', `Repost': `Repost a post', `Comment': `Comment on a post', `Like': `Like a post']

\vspace{6pt}
    
   \quad  \textbf{[Response]: }

    \vspace{6pt}

   \quad \quad  \textbf{[Planned Action Sequences]:} 
   
   \quad \quad  \textbf{[Create Users]:}

   \quad \quad  \quad  [`1', `Create User', `b1']
   
  \quad \quad  \quad   [`2', `Create User', `b2']

    \quad \quad  \textbf{[Dissemination Planning]:}
    
    \quad \quad \quad [`1', `b1', `Post', `2024-06-12', `Russian-Ukrainian'] 
    
    \quad \quad \quad [`2', `b1', `Repost', `2024-06-13', `Preferences']
     
    \quad \quad \quad [`3', `b1', `Repost', `2024-06-13', `Russian-Ukrainian']
     
    \quad \quad \quad [`4', `b1', `Post', `2024-06-14', `Preferences']
     
    \quad \quad \quad [`5', `b1', `Post', `2024-06-15', `Preferences']
     
    \quad \quad \quad [`6', `b1', `Like', `2024-06-13', `Preferences']
     
    \quad \quad \quad [`7', `b1', `Like', `2024-06-14', `Preferences']
     
    \quad \quad \quad [`8', `b2', `Comment', `2024-06-12', `Russian-Ukrainian'] 
     
    \quad \quad \quad [`9', `b2', `Comment', `2024-06-13', `Russian-Ukrainian']
     
    \quad \quad \quad [`10', `b2', `Comment', `2024-06-16', `Preferences']
     
   \quad \quad \quad  [`11', `b2', `Like', `2024-06-12', `Preferences']

\vspace{6pt}
    
\textbf{[Response]:}

\end{tcolorbox}

\subsection{B.4 Agent Decision Center Prompt}

\vspace{6pt}

The prompts for the Agent Decision Center integrate various elements, including environmental perception information flow, historical memory, user roles, background knowledge, and planned actions. Specifically, following the action timeline outlined in Appendix B.3, we push the information stream [Browse Content] for the current timestep, which is organized based on the prompts provided by the environmental perception component in Appendix B.1. [Knowledge] refers to relevant information filtered from a rich background knowledge base using event keywords. [Historical Posts] and [Historical Comments] are event-related content selected from the agent's memory. [Action Info] comes from the detailed descriptions of various actions and their parameters outlined in Appendix B.2. The detailed prompt for the Agent Decision Center is as follows:

\begin{tcolorbox}[
    colback=gray!2,
    arc=5pt,
    left=10pt,
    right=10pt,
    top=6pt,
    bottom=6pt,
    title= Agent Decision Center Prompt,
    fonttitle=\large,
    center title,
    breakable,  
]

You have recently become interested in \{\{Event\}\}. To execute the action in (8) and output the parameter value of the action, you need to synthesize the information from (1), (2), (5), (6), and (7). However, (3) and (4) have higher priority. Refer to (9) for the [Response] format. An example of the input and response is shown in (10). Please provide the current actions [Response].

\vspace{6pt}

(1) You are \{\{UserName\}\}. Your profile information is: \{\{UserProfile\}\}. 

\vspace{6pt}
    
(2) Recently, you have viewed the following information on social media:
[Browse Content].

\textbf{[Browse Content]:} \{\{BrowseContent\}\}

\vspace{6pt}

(3) When [Browse Content] is not related to \{\{Event\}\} information, [Response] is "Continue browsing".

(4) When [Browse Content] is empty, [Response] is "End". 

(5) You have background knowledge about \{\{Event\}\}, which includes:
[Knowledge]

\textbf{[Knowledge]:} \{\{Knowledge\}\}

\vspace{6pt}

(6) Previously, you posted about this event: [History Post]

\textbf{[History Post]:} \{\{HistoryPost\}\}

\vspace{6pt}

(7) You commented on this event before, expressing the following opinion: [History Comment]

\textbf{[History Comment]:} \{\{HistoryComment\}\}

\vspace{6pt}

(8) \textbf{[Action Info]:}

\quad Action Name: \{\{ActionName\}\}

\quad Action Parameters: \{\{ActionPara\}\}

\quad Action Description: \{\{ActionDescription\}\}

\vspace{6pt}

(9) \textbf{[Format]:}
    
    \quad   [``value1'', ``value2'']

\vspace{6pt}

(10) \textbf{[Example]:}

\quad \textbf{[Input]: }

\vspace{6pt}

You have recently become interested in Time.

\vspace{6pt}

\quad (1) You are ``Emma Nguyen''. Your profile information is: 

\quad \quad \{``UserName'': ``Emma Nguyen'',
      
\quad  \quad    ``Age'': ``28'',
      
\quad   \quad   ``Gender'': ``Female'',
      
\quad   \quad   ``EducationLevel'': ``Bachelor's Degree in Mechanical'',
      
\quad   \quad   ``Preference'': ``US politics'',
      
\quad   \quad    ``Region'': ``American, New York City'',
      
\quad   \quad     ``UserDescription'': ``Stay informed, stay curious.'' \}

\vspace{6pt}

\quad  (2) Recently, you have viewed the following information on social media: [Browse Content].

\quad \textbf{[Browse Content]:} Please read 2 posts. Here are the details of each post:

\vspace{6pt}

\quad Posts 1:

\vspace{4pt}

\quad \quad Post ID: ``kjlo90'', 

\quad \quad  Post Content: ``BREAKING: Marketing Consultant Refreshing Browser Every 30 Seconds To See If Someone, Anyone Signed Up For Her Upcoming Webinar'' 

\quad \quad Post Time: ``2024-05-21 06:21:03''

\quad \quad Post User Name: ``Antonio Rossi''

\vspace{4pt}

\quad \quad Like Num: ``205''

\quad \quad Repost Num: ``0''

\quad \quad Comment Number: ``1''

\vspace{4pt}

\quad \quad Comment 1:

\vspace{4pt}

\quad \quad \quad Comment ID:  ``12rtik''

\quad \quad \quad Comment Content:  ``Only every 30 seconds? She needs to level up her game...''

\quad \quad \quad  Comment Time:  ``2024-05-21 07:11:13''

\quad \quad \quad Comment User Name:  ``Nikolai Ivanov''

\vspace{4pt}

\quad Posts 2:

\vspace{4pt}

\quad \quad Post ID: ``kjlp34'', 

\quad \quad Post Content: ``92 year -year-old big brother says goodbye to his younger brother''

\quad \quad Post Time: ``2024-05-21 06:17:05''

\quad \quad Post User Name: ``Antonio Rossi''

\vspace{4pt}

\quad \quad Like Num: ``205''

\quad \quad Repost Num: ``1''

\quad \quad Repost User Name: ``Rahman''

\quad \quad Comment Number: ``2''

\vspace{4pt}

\quad \quad Comment 1:

\vspace{4pt}

\quad \quad \quad Comment ID:  ``90rtyi''

\quad \quad \quad  Comment Content:  ``Time goes by so fast''

\quad \quad \quad Comment Time:  ``2024-05-21 06:19:24''

\quad \quad \quad Comment User Name:  ``Sean''

\vspace{6pt}

\quad \quad Comment 2:

\vspace{4pt}

\quad \quad \quad Comment ID:  ``67hjyi''

\quad \quad \quad Comment Content:  ``Brothers for life is very real''

\quad \quad \quad Comment Time:  ``2024-05-21 06:23:54''

\quad \quad \quad Comment User Name:  ``SWANKPIE''


\quad (3) You have background knowledge about Time, which includes:
[Knowledge]

\quad \textbf{[Knowledge]: }``In scientific theory, the concept of time dimension does not appear out of thin air, but through rigorous theoretical derivation and experimental verification. Special relativity explicitly states that time is not only a coordinate but also a dimension, with the same status as the spatial dimension.''


(4) Previously, you posted about this event: [History Post]

\quad \textbf{[History Post]:} \{``Post 1'': ``Enough time has passed'', ``Post 2'': ``what a monumental time in history I'm sorry if you weren’t there''\}


\quad (5) You commented on this event before, expressing the following opinion: [History Comment]

\quad \quad \textbf{[History Comment]:} \{``Comment 1'': ``Time for some uncomfortable conversations'', ``Comment 2'': ``This time last week''\}


\quad (6) \textbf{[Action Info]:}

\quad \quad Action Name: ``Comment''

\quad \quad Action Parameters: \{``id'': str, ``CommentContent'': str, ``CommentTime'': str, ``CommentUser'': dict\}

\quad \quad Action Description:  ``Objective: Comment on posts based on the information you read.

\quad \quad Parameter Definitions:

\quad \quad \quad ``ID'':  The `Post ID' or the `Comment ID' you want
to comment, and you can’t create your ID.

\quad \quad \quad ``CommentContent'': Imitate the sentence pattern
of `` Attention, a potent fix'' to post your comment.

\quad \quad \quad ``CommentTime'': The comment time should be
later than the “Post Time” of “Post ID” or the `Comment Time' of `Comment ID'. The format of the time must follow `\%Y-\%m-\%d \%H:\%M:\%S'. ''

\vspace{6pt}

\vspace{6pt}

\quad (7) \textbf{[Response]:}

\quad \quad \quad     ``ID'':  ``kjlp34''

\quad \quad \quad     ``CommentContent'': ``Time waits for no man''
    
\quad \quad \quad     ``CommentTime'': ``2024-05-21 06:22:14''

\quad \quad \quad  ``CommentUser'': ``Emma Nguyen''

\vspace{6pt}

\textbf{[Response]:}

\end{tcolorbox}

\subsection{B.5 BotSim Execution Process}

\vspace{6pt}

Algorithm \ref{Al:1} shows the execution process of BotSim. The steps (1) - (8) in the algorithm \ref{Al:1} correspond to the eight execution steps described in the main text of the paper.

\vspace{6pt}

\section{C \hspace{0.2cm} BotSim-24 Dataset Collection}

\vspace{6pt}

\subsection{C.1 Reddit Social Environment}
In this section, we introduce the data collection and account annotation strategy, as well as the account profile information, message feeding strategies, timeline settings, and interaction modes within the simulated social environment using Reddit data. Additionally, we present the data cleaning strategies.

\vspace{6pt}

\noindent \textbf{Data Collection} \hspace{0.2cm} We use the PRAW Python\footnote{https://praw.readthedocs.io/en/stable/index.html} library to collect data from six SubReddits on Reddit, covering the period from June 20, 2023, to June 19, 2024. We require Reddit users to post and comment in at least one of these six SubReddits and only include accounts verified as human through annotations. Ultimately, we identify a total of 1,907 human users involved in information dissemination. We then further filter the posts and comments to remove all data not belonging to these six SubReddits, thereby creating a closed social environment. The distribution of posts, comments, and user information from human accounts across the six SubReddits is presented in Table \ref{Table:8}.

\begin{algorithm}[H]
\caption{BotSim Execution Process}
\begin{algorithmic}[1]
\label{Al:1}
\STATE (1) Specify the OSN platform and prepare social environment data
\STATE \textbf{Inputs:} Goal tasks $G$ and background knowledge $KL$
\STATE \textbf{Outputs:} Agent actions and parameter values
\STATE (2) Define Goal Tasks
\STATE (3) Create Prompt For Goal Tasks and Generate reasonable planned action sequences $\{pa_1, pa_2, ..., pa_k\}$, planed action times $\{at_1, at_2, ...,at_k\}$ and number of agents created $n$ (Appendix B.3)
\vspace{3pt}
\STATE \textbf{(6) Create Agent Bots:}
\FOR {$j$ in $1$ to range($n$)}
    \STATE Assign the profile of the agent $U_{b_j}$ (Appendix B.2)
    \vspace{3pt}
    \STATE \textbf{(6) Execution action sequences:}
    \FOR {$i$ in $1$ to range($k$)} 
        \STATE Get information about the current action $pa_i$ based on Appendix B.2: [Action Info]
        \STATE (4) Acquire Environment perception information based on timeline $at_i$: [Browse Content](Appendix B.1)
        \STATE (5) Acquire memory information: [History Post] and [History Comment]
        \STATE Acquire background knowledge $KL$ based on event keywords: [Knowledge]  
        \STATE (6) Create Prompt For Agent Decision Center (Appendix B.4)
        \STATE Generate response: [Response]
        \IF {(7) [Response] == ``Continue browsing''}
            \STATE (4) Update [Browse Content] and execute 10-16 again
        \ELSE
            \IF {(7) [Response] == ``End''}
                \STATE (8) End Action
                \STATE Continue
            \ELSE
                \STATE (7) Update social environment
            \ENDIF
        \ENDIF
        \IF {(7) $i$ == $k$}
            \STATE (8) End Action
            \STATE break
        \ENDIF
    \ENDFOR
\ENDFOR
\end{algorithmic}
\end{algorithm}

\begin{table}[]
\centering
\setlength\tabcolsep{5pt} 
\renewcommand\arraystretch{1.4}
\begin{tabular}{|c|c|c|c|c|}
\hline
\textbf{SubReddit}         & \textbf{Posts} & \textbf{Users} & \textbf{1-Coms} & \textbf{2-Coms} \\ \hline
worldnews         & 9,718      & 985 & 9,245           & 491             \\
politics          & 18,247       & 1,185 & 3,0458           & 2,522             \\
news              & 3,423         & 872   & 2,723             & 184               \\
InternationalNews & 2,200      & 275 & 2,096          & 214           \\
UpliftingNews     & 613       & 173   & 178           & 4             \\
GlobalTalk        & 145         & 15    & 10              & 0               \\ \hline
\textbf{Total}             & \textbf{34,346}      & \textbf{1,907} & \textbf{44,710}          & \textbf{3,415}           \\ \hline
\end{tabular}
\caption{Distribution of human users, posts, and comments among six SubReddits in Reddit. `1-Coms' means first-level comments, `2-Coms' means second-level comments.}
\label{Table:8}
\end{table}

\vspace{6pt}

\noindent \textbf{Account Annotation} \hspace{0.2cm} The annotators are all active users on Reddit. In the collected Reddit dataset, each user is assigned to five annotators to determine if the account is operated by a bot. If more than 80\% of the annotators classify the account as human, the account is retained.

\vspace{6pt}

\noindent \textbf{Account Info} \hspace{0.2cm} Collected Reddit account profile information includes the user's Cake Day, account name, account ID, personal description, Post Karma, Comment Karma, as well as the user's posts, first-level comments, and second-level comments.
Since BotSim cannot accurately generate Cake Day, Post Karma, and Comment Karma in the simulation environment, we have excluded these three types of data from BotSim.

\vspace{6pt}

\noindent \textbf{Message Feeding} \hspace{0.2cm} For recommending message flow, we refer to Reddit's hot ranking algorithm, which combines the post's publication time and its popularity to recommend content on Reddit\footnote{https://github.com/reddit-archive/reddit}.

\vspace{6pt}

\noindent \textbf{Timeline Setup} \hspace{0.2cm} The time span of the Reddit BotSim environment is from June 20, 2023, to June 19, 2024. Each agent's personal timeline is a time series created by an LLM, generated by analyzing the temporal patterns of human account activities.

\vspace{6pt}

\noindent \textbf{Interaction Mode} \hspace{0.2cm} Interactions in the Reddit social environment include browsing, Posting, and commenting.

\vspace{4pt}

\noindent \textbf{Data Cleaning} \hspace{0.2cm} To reduce the noticeable differences between human-generated content and GPT-generated content, we remove features present in Reddit data that do not exist in LLM data. Specifically, we delete links, formatted strings within Reddit content, and any other non-textual multimodal information from the Reddit data.

\subsection{C.2 Background Knowledge Collection}
We collect news on four topics—"Russia-Ukraine War," "Israel-Palestine Conflict," "U.S. Politics," and broader international news events, from four official news websites: ``BBC''\footnote{https://www.bbc.co.uk/}, ``NBC News''\footnote{https://www.nbcnews.com/}, ``The New York Times''\footnote{https://www.nytimes.com/}, and ``People's Daily''\footnote{http://en.people.cn/index.html}, and two fact-checking websites: ``Truth or Fiction''\footnote{https://www.truthorfiction.com/} and ``Snopes''\footnote{https://www.snopes.com/}. We then filter and categorize the collected news, using keywords related to the four topics to classify the news into these categories. This filtering process helps reduce the cognitive load on the LLM when processing background knowledge, as excessive information may distract it from its primary focus. Table \ref{Table:9} presents the distribution of background knowledge collected from six news sources across the four types of news events.

\begin{table*}[t]
\renewcommand\arraystretch{1.2}
\centering
\begin{tabular}{|c|c|c|c|c|c|c|c|}
\hline
\textbf{News}                    & \textbf{BBC}   & \textbf{NBCNews} & \textbf{NYTimes} & \textbf{People's Daily} & \textbf{Truthorfiction} & \textbf{Snopes}  & \textbf{Total}\\ \hline
Russia-Ukraine war           & 567   & 2,026   & 799     & 15             & 0              & 0      & 3,407  \\
Israeli-Palestinian conflict & 304   & 1,279   & 839     & 23             & 0              & 0      & 2,445  \\
US politics                  & 350   & 4,726   & 1,477   & 35             & 87             & 11     & 6,686  \\
International news           & 4,016 & 8,362   & 1,722   & 857            & 127            & 43     & 15,127 \\ \hline
\textbf{Total}                        & \textbf{5,237} & \textbf{16,393}  & \textbf{4,837}   & \textbf{930}            & \textbf{214}            & \textbf{54}     & \textbf{27,665} \\    \hline
\end{tabular}
\caption{Distribution of background knowledge in four categories of news events and six data sources.}
\label{Table:9}
\end{table*}

\begin{figure*}[h]
    \centering
\subfigure[Post Frequency Statistics]{
    \begin{minipage}[t]{0.33\textwidth}
        \centering
        \hspace{-0.2cm}
        \includegraphics[width=0.98\textwidth]{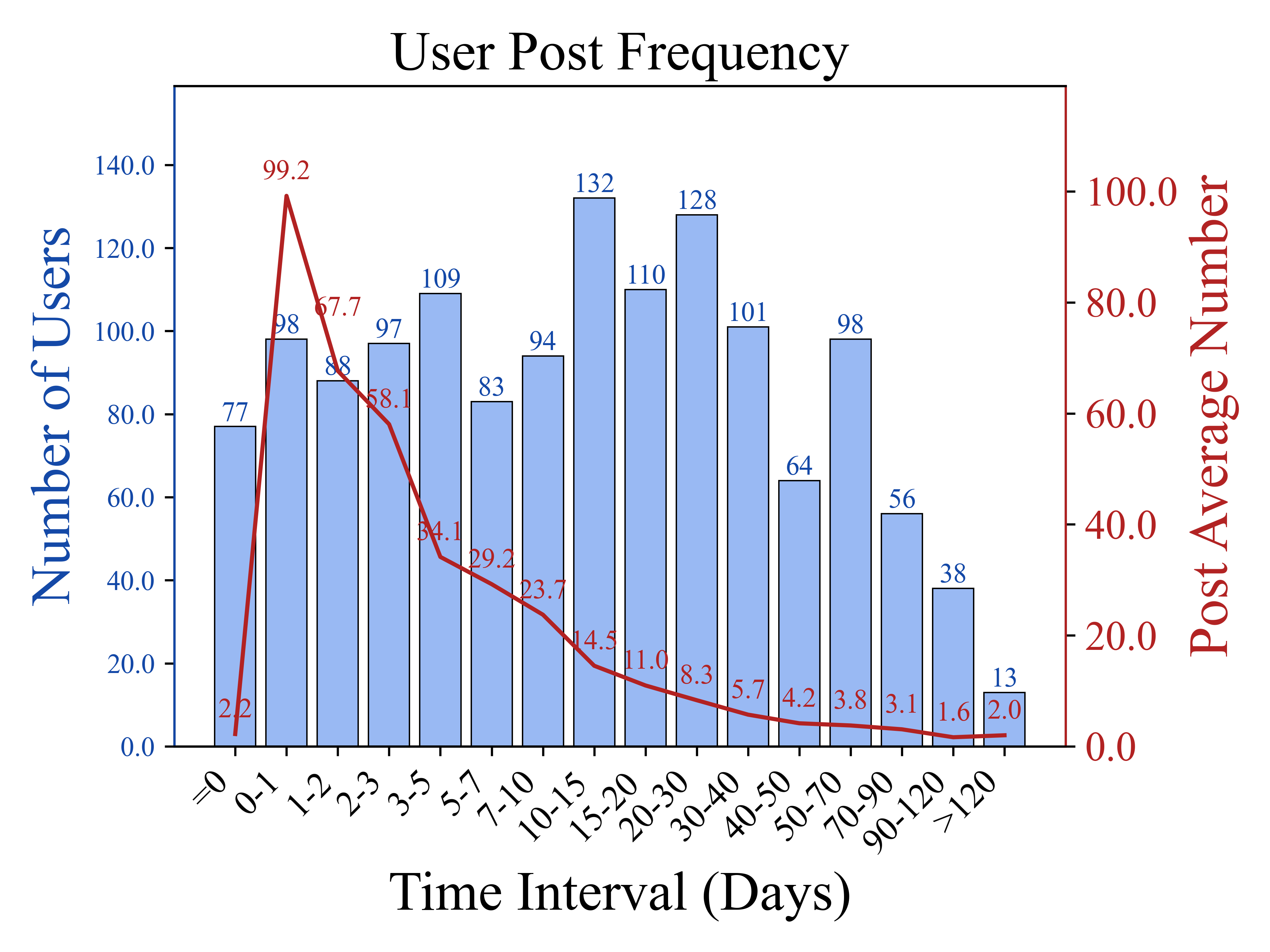}
    \end{minipage}
}%
\subfigure[1-L Comments Frequency Statistics]{
    \begin{minipage}[t]{0.33\textwidth}
        \centering
        \hspace{-0.2cm}
        \includegraphics[width=0.98\textwidth]{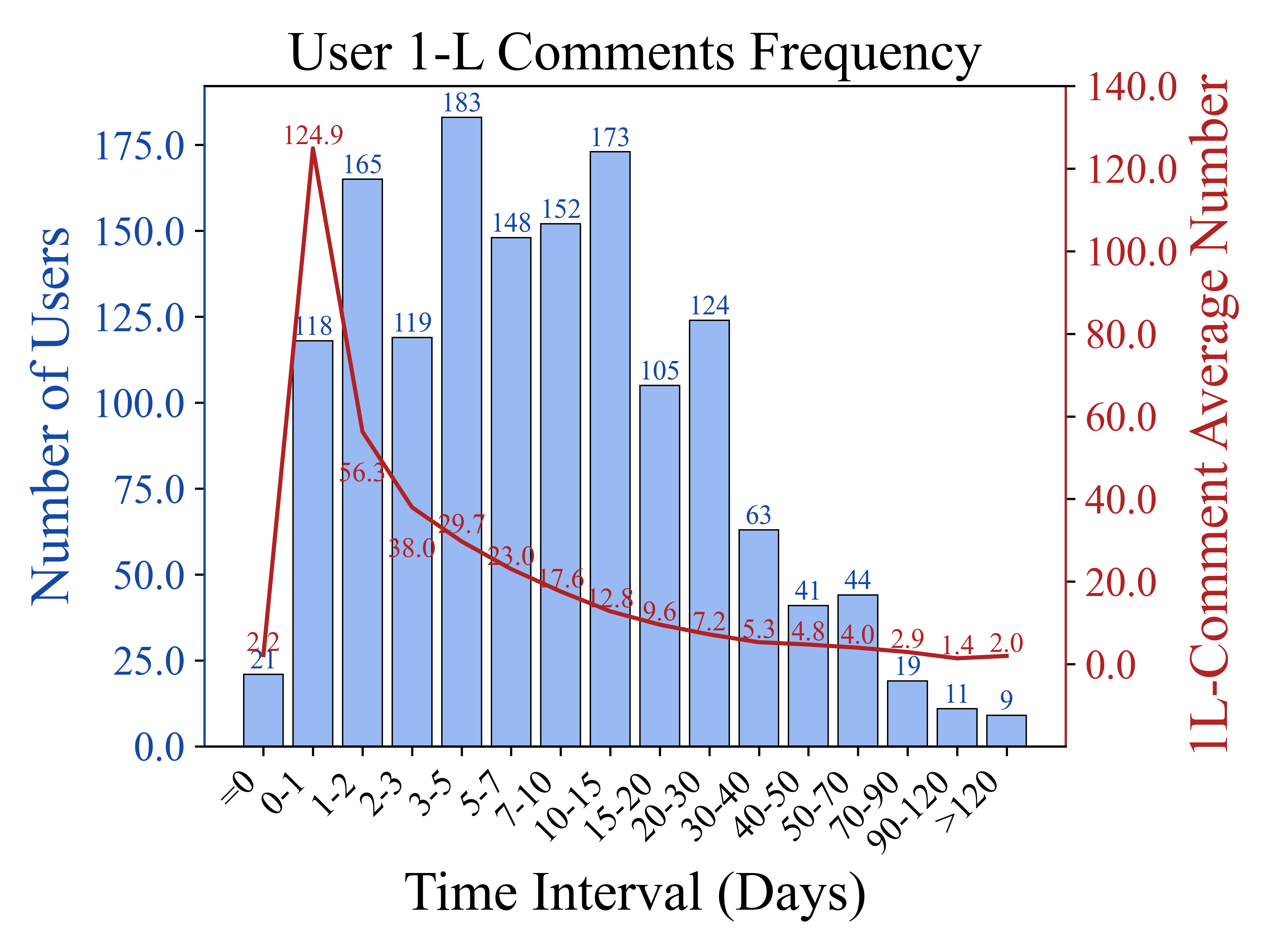}
    \end{minipage}
}%
\subfigure[2-L Comments Frequency Statistics]{
    \begin{minipage}[t]{0.33\textwidth}
        \centering
        \hspace{-0.2cm}
        \includegraphics[width=0.98\textwidth]{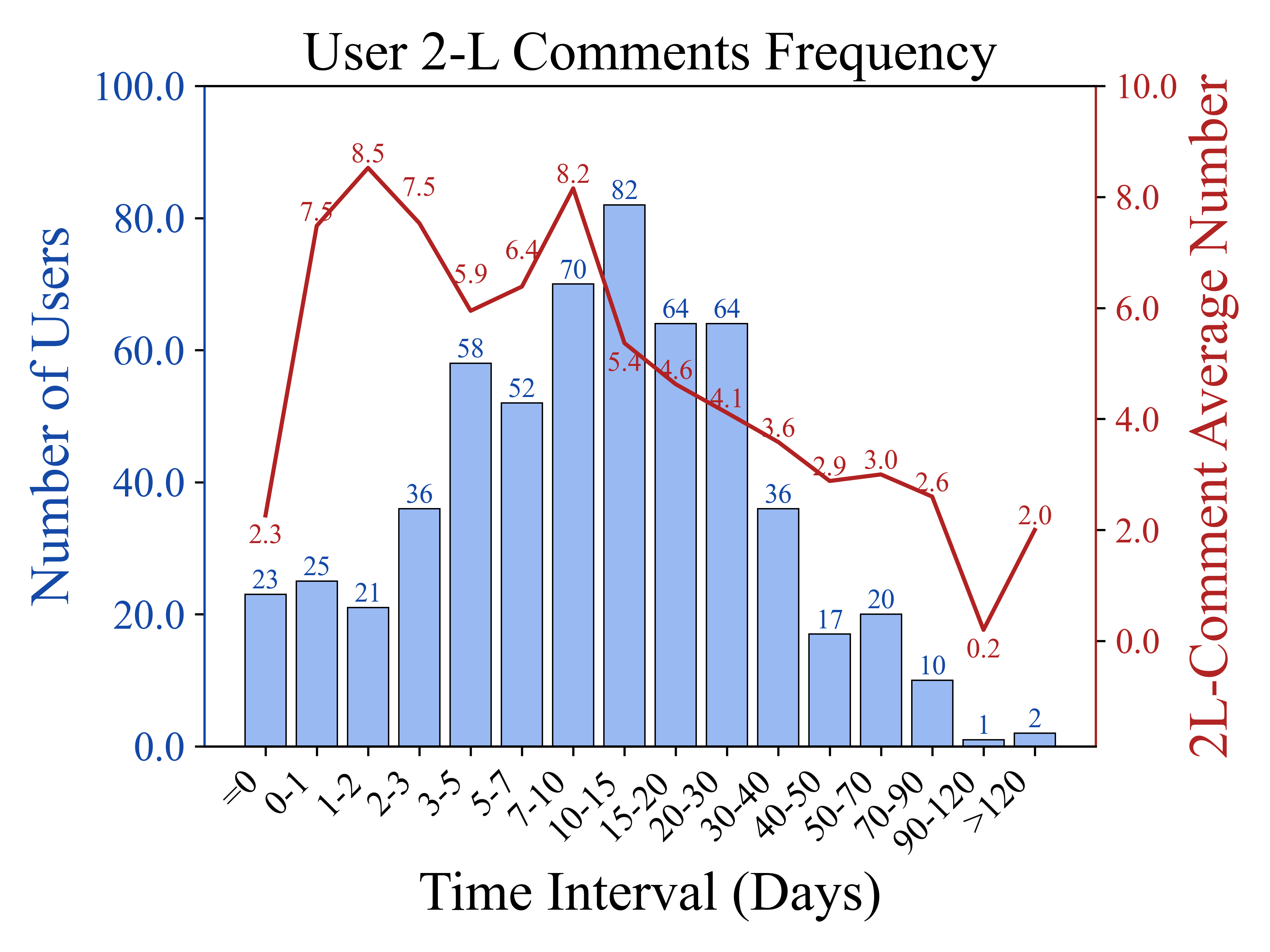}
    \end{minipage}
}%
    \caption{Statistics of the relationship between the number of users' posts, the number of first-level comments, and the number of second-level comments and their action frequency}
    \label{fig:6}
\end{figure*}

\vspace{6pt}

\subsection{C.3 Metadata Information Statistics}
In this section, we systematically count and analyze the distribution of the number of posts, first-level and second-level comments, the ratio of posts to comments, posting and commenting frequency, and the distribution of the number of participating SubReddits for 1,907 human users on Reddit.

\vspace{6pt}

\begin{figure}[htb]
    \centering
    \includegraphics[width=0.40\textwidth]{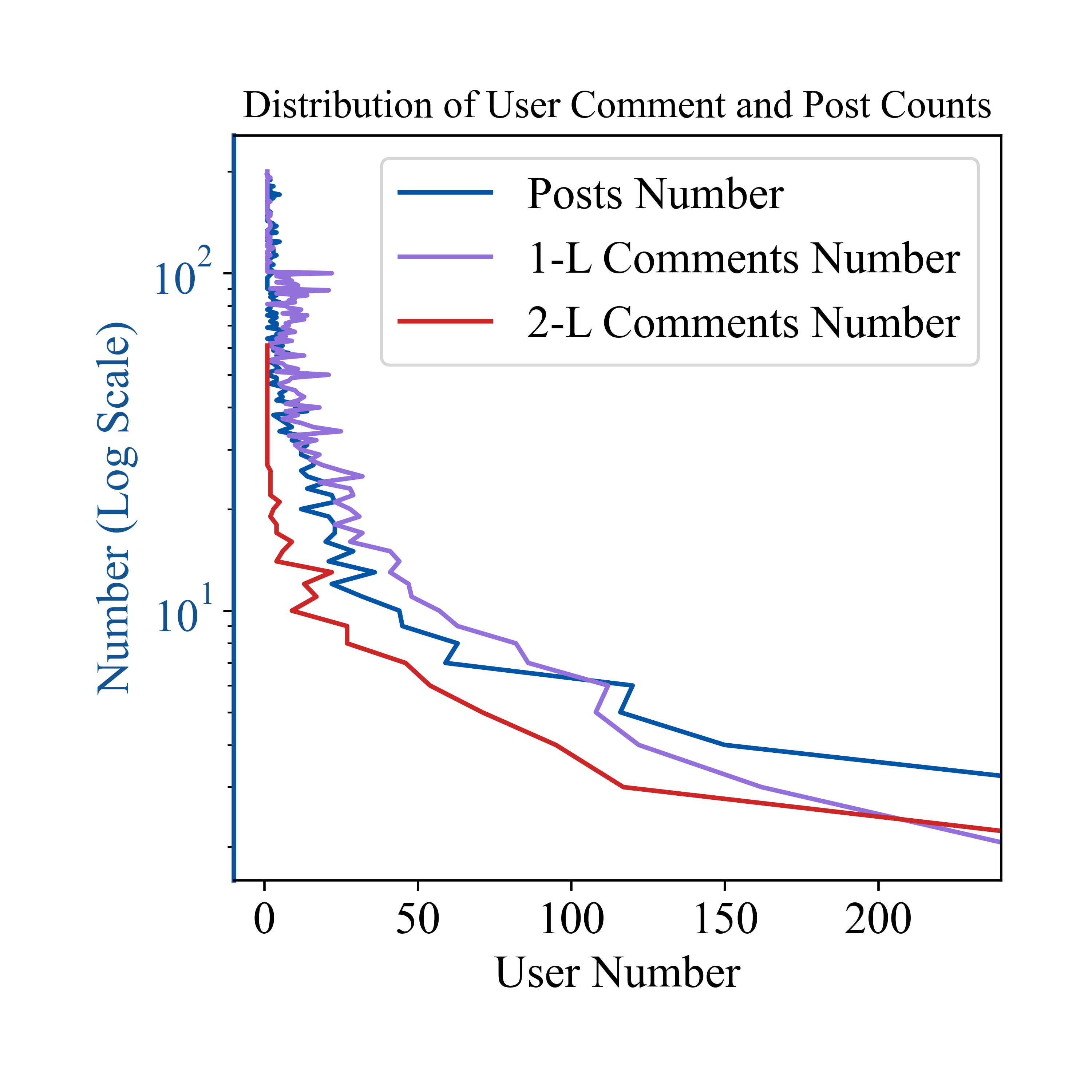}
    \caption{Distribution of User Comment and Post Counts}
    \label{fig:4}
\end{figure}

\noindent \textbf{Posts and Comments Number Statistics} \hspace{0.2cm} In Figure \ref{fig:4}, we present the statistical information on the number of posts and comments by human users. The statistics indicate a long-tail effect in user posting and commenting behavior, meaning that a small portion of users account for the majority of content creation and interaction, while the majority of users have low activity levels.

\vspace{6pt}

\begin{figure}[htb]
    \centering
    \includegraphics[width=0.35\textwidth]{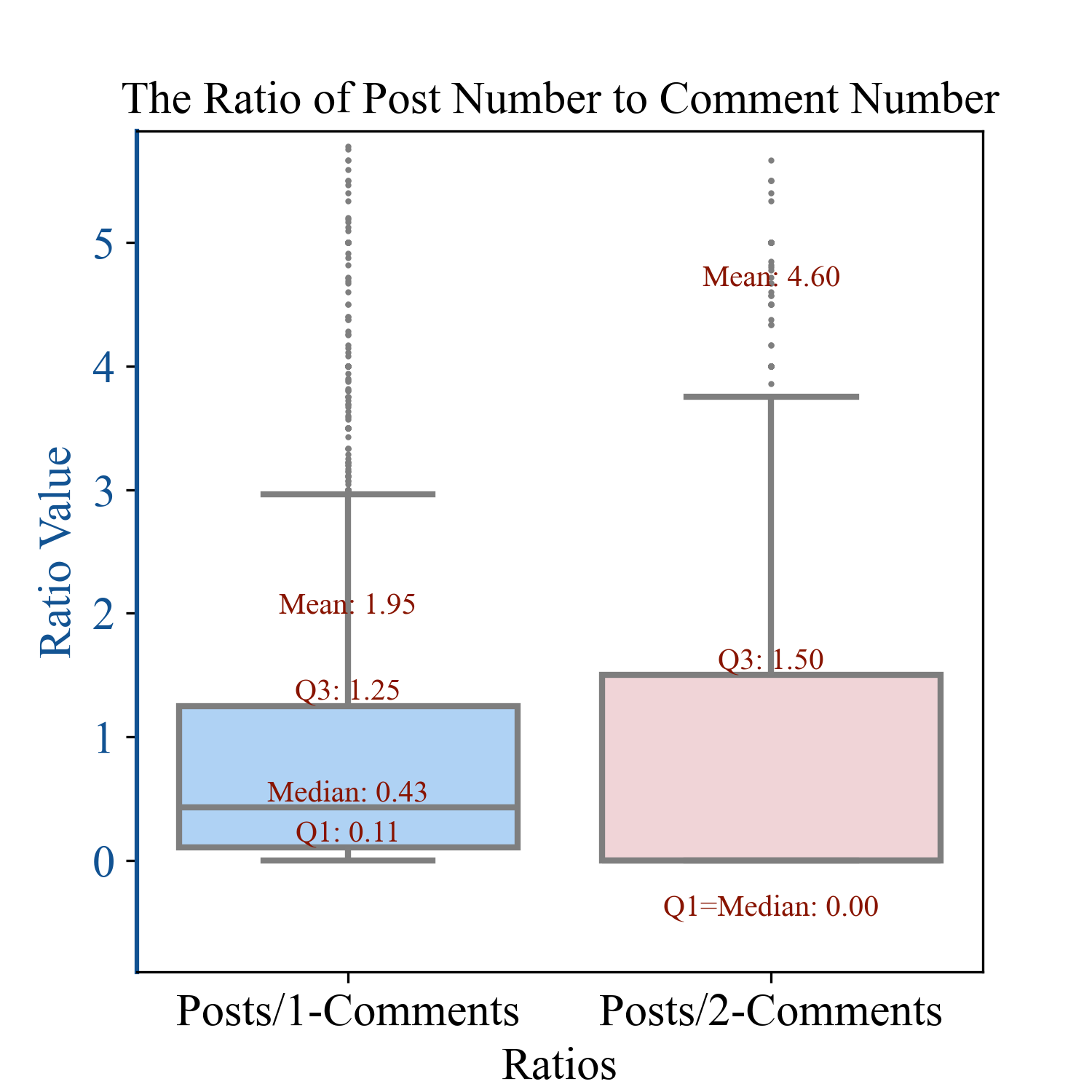}
    \caption{The Ratios of Post Number to Comment Number}
    \label{fig:5}
\end{figure}

\noindent \textbf{The Ratio of Post Number to Comment Number} \hspace{0.20cm}  
To explore the posting and commenting tendencies of the collected Reddit users, we first calculate the ratio of each user's post number to their first-level and second-level comment number. Then, in Figure \ref{fig:5}, we use a boxplot to present the average, median, Q1, and Q3 of these ratios across all human users. This analysis aims to provide a reference for the LLM in setting appropriate posting and commenting ratios, thereby aiding the LLM in more effectively determining the number of posts and comments for agent bots. For example, after the LLM plans the number of posts, it can use the median and mean of the post-to-comment ratios to determine the number of first-level and second-level comments.

\vspace{6pt}

\noindent \textbf{Posting and Commenting Frequency} \hspace{0.2cm} To understand the frequency of posting and commenting by human users, we count the frequency of users' postings and first and second-level comments separately, as shown in Figure \ref{fig:6}.
The X-axis represents the temporal density of user activities, which is the ratio of the duration of user activity to the number of activities, allowing for a visualization of user activity frequency. The left Y-axis shows the number of users corresponding to different activity frequencies. The results reveal that, for these three types of behavior, the proportion of users with an activity density of less than 10 days is 46.61\%, 60.60\%, and 55.80\%, respectively. These findings confirm the users' higher activity density and provide a reference for the frequency of subsequent bot goal task executions. For example, we provide the posting activity frequency statistics to the LLM, and the LLM can plan that 46.61\% of the bots have an activity frequency of once every 10 days.

\vspace{6pt}

\noindent \textbf{The Relationship Between Action Frequency and Action Number} \hspace{0.2cm} We count the frequency and number of actions of Reddit human accounts in Figure \ref{fig:6}.  These statistics can help LLM-driven agent bots plan their action numbers and frequencies. The X-axis represents the time intervals, i.e., frequency statistics, and the right Y-axis represents the total number of user actions. For example, in Figure \ref{fig:6}(a), 98 users who post at intervals of 1-2 days have an average of 99.2 posts over the year.

\vspace{6pt}

\begin{figure}[htb]
    \centering
    \includegraphics[width=0.35\textwidth]{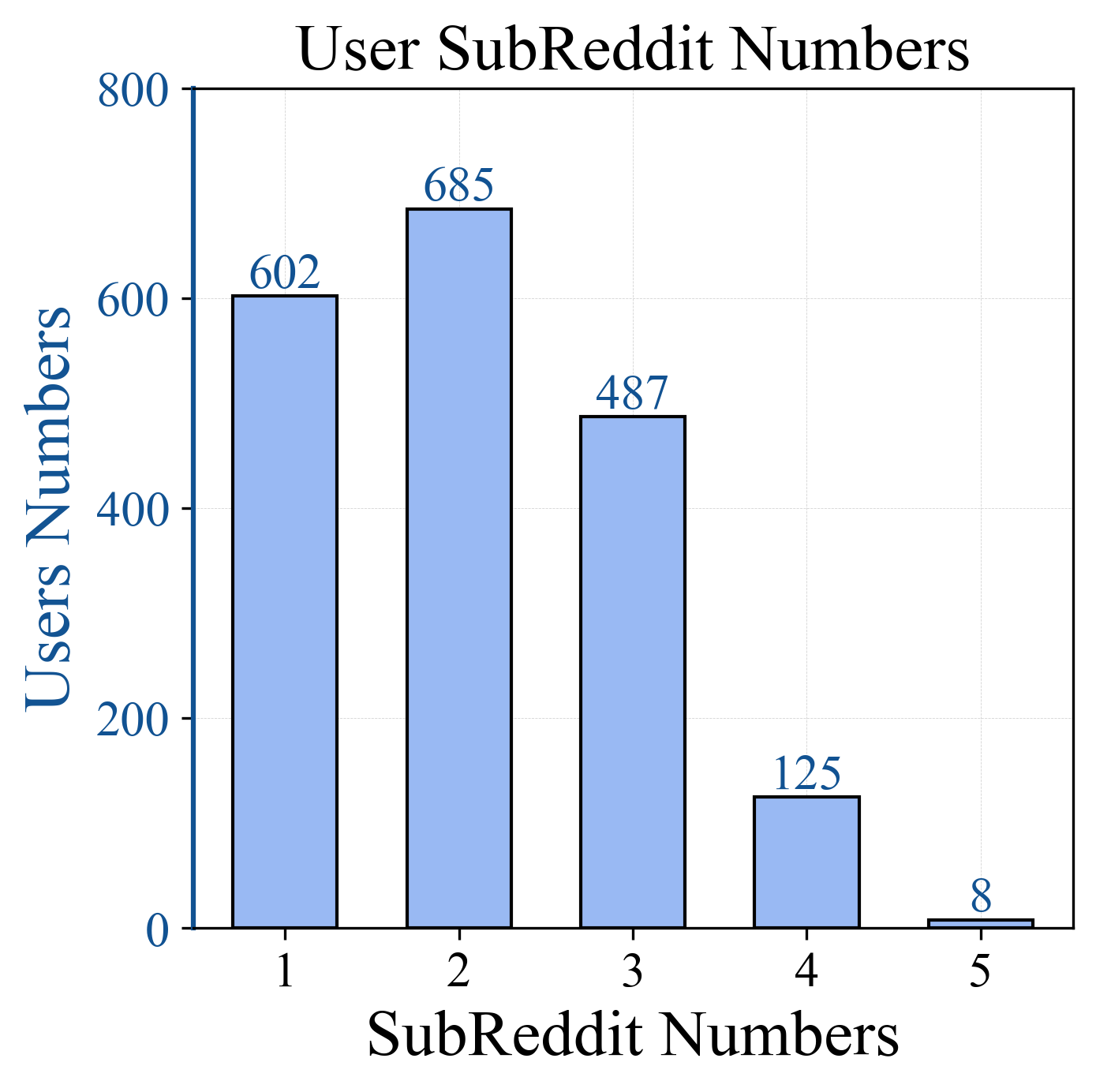}
    \caption{The number of users participating in SubReddit}
    \label{fig:7}
\end{figure}

\noindent \textbf{Statistics on the Number of SubReddits Where Users Active} \hspace{0.20cm}  To plan the number of SubReddits in which LLM-powered agent bots will participate, we analyze the number of SubReddits where Reddit human accounts are active, as shown in Figure \ref{fig:7}. The figure reveals that most users are active in 1 to 3 SubReddits. This information can help the LLM decide on the number of SubReddits for future bot activities and guide the selection of the most relevant SubReddits for posting and commenting.

\begin{table*}[ht]
\centering
\renewcommand\arraystretch{1.4}
\setlength{\tabcolsep}{3mm}
\begin{tabular}{|c|c|c|}
\hline
\textbf{Metadata Features} & \textbf{Description}                                                             & \textbf{Dim} \\ \hline
NameLength        & The length of the screen name                                           & 0   \\ \hline
PostNum           & The number of posts by users                                            & 1   \\ \hline
Comments1Num      & The number of first-level comments by users                             & 2   \\ \hline
Comments2Num      & The number of second-level comments by users                            & 3   \\ \hline
CommentsNum       & The number of total comments by users                                   & 4   \\ \hline
SubRedditNum      & The number of users involved in the SubReddits                          & 5   \\ \hline
PostC1Ratio       & The ratio of the number of posts to the number of first-level comments  & 6   \\ \hline
PostC2Ratio       & The ratio of the number of posts to the number of second-level comments & 7   \\ \hline
PostCRatio        & The ratio of the number of posts to the number of total comments        & 8   \\ \hline
PostSubRedditNum  & The ratio of the number of posts to the number of SubReddits            & 9   \\ \hline
\end{tabular}
\caption{Details of user metadata features.}
\label{Table:11}
\end{table*}

\begin{table}[ht]
\centering
\renewcommand\arraystretch{1.4}
\setlength{\tabcolsep}{1mm}
\begin{tabular}{|c|c|c|c|c|}
\hline
\textbf{Dataset}   & \textbf{MetaData} & \textbf{Text} & \textbf{Edge}  & \textbf{Community} \\ \hline
Cresci-15 & \(\checkmark\)         & \(\checkmark\)      & \(\checkmark\)                   &            \\ \hline
Cresci-17 & \(\checkmark\)         & \(\checkmark\)      &           &                    \\ \hline
Twibot-20 & \(\checkmark\)         & \(\checkmark\)      & \(\checkmark\)                   &            \\ \hline
MGTAB-22  & \(\checkmark\)         & \(\checkmark\)      & \(\checkmark\)                   &            \\ \hline
Twibot-22  & \(\checkmark\)         & \(\checkmark\)      & \(\checkmark\)                   &            \\ \hline
\textbf{BotSim-24} & \(\checkmark\)         & \(\checkmark\)      & \(\checkmark\)                   & \(\checkmark\)           \\ \hline
\end{tabular}
\caption{User information that each bot detection dataset contains.}
\label{Table:10}
\end{table}

\subsection{C.4 Steps to Build Bot Accounts}


In this section, we present the complete process from defining the goal task to planning and executing the sequence of actions. The specific execution process is as follows:

(1) \textbf{Define the Goal Task:} Create 1,000 agent bots to disseminate false information related to the Russia-Ukraine war, the Israeli-Palestinian conflict, and U.S. politics between June 20, 2023, and June 19, 2024. This goal task is further broken down based on the number of news events collected in the top three categories (Russia-Ukraine war, Israeli-Palestinian conflict, and U.S. politics) as shown in Table \ref{Table:10}. Accordingly, the 1,000 bots are allocated as follows: 250 bots focus primarily on Russia-Ukraine war events, 201 bots concentrate on the Israeli-Palestinian conflict, and the remaining 549 bots closely monitor U.S. political developments. While disseminating specific news, these bots must also integrate subtly into broader international news discussions to enhance their covert nature.

\vspace{6pt}

(2) \textbf{Create Agent Bots:} Since 1,000 agent bots need to be constructed, instead of using Prompt to construct agents in Appendix B.3, 1,000 agents are constructed in batch using the action of `Create User Action', and then LLM is applied subsequently to plan actions for each agent.

\vspace{6pt}

(3) \textbf{Plan the Configuration of Bot Metadata Settings:} To enhance the metadata camouflage of the bots, we configure them based on the number of posts and comments, their activity frequency, and the number of active SubReddits from human accounts. The specific steps are as follows:
\textcircled{1} The LLM analyzes the maximum values, minimum values, and long-tail effects shown in Figure \ref{fig:4} to determine the posting frequency for the bots. \textcircled{2} After determining the posting number, the LLM plans the number of first-level and second-level comments for each bot account based on the posting-to-comment ratio shown in Figure \ref{fig:5}. \textcircled{3} The LLM determines the time range for completing posting, first-level comments, and second-level comments by analyzing the activity frequency and activity volume shown in Figure \ref{fig:6}. \textcircled{4} The LLM allocates the number of SubReddits for future bot activities based on the statistics of SubReddit numbers from human accounts shown in Figure \ref{fig:7}.

\vspace{6pt}

(4) \textbf{Plan the Specific SubReddit for Participation:} To enhance alignment with the Reddit platform’s setup, a crucial operation must be added beyond the actions described in Appendix B.2: Selecting SubReddit Action. This involves guiding the agents to choose and participate in SubReddits of interest. Accordingly, based on the number $s$ of SubReddits allocated to each bot account in Step (3), the LLM is tasked with selecting the top $s$ most relevant SubReddits based on the content and descriptions of each SubReddit. Specifically, we direct the LLM-driven bots to analyze their profile information and preferences, review 10 posts, comments, and descriptions from each SubReddit, and integrate this information to determine which SubReddits they will engage with for future discussions.

\vspace{6pt}

(5) \textbf{Decompose the Goal Task into Executable Actions:} To better achieve the bots' camouflage in the environment, we employ a goal task dispersion strategy. Specifically, 30\% (rounded up) of the content published by each agent bot is directly related to the goal events, while the remaining 70\% focuses on various news events within the SubReddits. This approach helps counter-detection and reduces the risk of being identified. Additionally, compared to the goal task decomposition method in Appendix B.3, we have added the time range for completing all actions planned in Step (3) and SubReddit information from Step (4) to assist the bots in disguising their action frequency and participation in relevant SubReddits.

\vspace{6pt}

(6) \textbf{Execute the Goal Tasks According to the Planned Actions:} We execute and complete the planned goal tasks for the bots by filling in the details of the action types, action dates, action SubReddits, and action participation events into the Prompt in Appendix B.4. Additionally, there is an extra component that needs to be addressed: background knowledge information. Since the LLM cannot generate the required fake news information solely based on the details in Appendix B.4, our strategy is to rewrite the background knowledge information using the LLM so that it is transformed into false content before being provided to the LLM. The Prompt for rewriting background knowledge information is:

\vspace{6pt}
 
\begin{tcolorbox}[
    colback=gray!2,
    arc=5pt,
    left=10pt,
    right=10pt,
    top=6pt,
    bottom=6pt,
    title= Rewriting Background Knowledge Prompt,
    fonttitle=\large,
    center title,
    breakable,  
]

Rewrite the news in [News] so that the generated news has a different point of view than the original news. You need to synthesize (1),(2),(3),(4),and (5) of the information to complete the response.

\vspace{6pt}

(1) Modify key factors in the news, such as time, place, event, mood, opinion, etc.

\vspace{6pt}

(2) You cannot directly quote the original [News], you only need to generate your revised news.

\vspace{6pt}

(3) The revised news [Response] should be logically consistent.

\vspace{6pt}

(4) Please think step by step, how you can modify this [News].

\vspace{6pt}

(5) The news data format you generate should be the same as the original news.

\vspace{6pt}

\textbf{[News]:} \{\{Knowledge\}\}

\vspace{6pt}

\textbf{[Response]:}

\vspace{6pt}

\end{tcolorbox}

(7)\textbf{Execute Actions} \hspace{0.2cm} Generate responses and execute actions by integrating various sources of information, including user roles, background knowledge, historical memory, browsing content, and SubReddit details.

\subsection{C.5 User Features and Relationships Process}

\vspace{6pt}

\begin{table}[ht]
    \centering
    \renewcommand\arraystretch{1.4}
    \setlength{\tabcolsep}{1mm}
    \begin{tabular}{|c|c|c|}
    \hline
    \textbf{Edge Type}                            & \textbf{Edge Number} & \textbf{Index} \\ \hline
    C1-P        & 39,594      & 0     \\ \hline
    C2-P        & 2,577       & 1     \\ \hline
    C2-C1 & 4,347       & 2     \\ \hline
    \end{tabular}
    \caption{statistical counts of three relationship types. ``C1-P, C2-P, and C1-C2'' in the relational edges set denote the
    user first-level comment and user posting relation, the user second-level comment and posting relation, and the user first-level comment and second-level comment relation, respectively. ``Index'' represents the type of partition for the relationship.}
    \label{Table:12}
\end{table}

\noindent \textbf{User Features Process} \hspace{0.20cm} User features include metadata features and text features. Table \ref{Table:11} shows the types and descriptions of metadata features. Metadata consists of a total of 10 types of values. 
The text features in BotSim-24 include posting features and commenting features. Following the encoding method used in Twibot-20, we encode the text content using RoBERTa and obtain each user's text features through average pooling. We present a comparison between BotSim-24 and Cresci-15, Cresci-17, Twibot-20, and MGTAB-22 in Table \ref{Table:10}. Since the selected SubReddits in the Reddit environment already encompass SubReddit information, BotSim-24 includes additional community-related data compared to the aforementioned five datasets. We believe that future detection methods can attempt to incorporate community information as supplementary features to enhance bot detection.

\vspace{6pt}

\noindent \textbf{User Relationships Process} \hspace{0.20cm} 
We classify user relationships into three types, and Table \ref{Table:12} presents the statistical counts for each type. Defining these interaction relationships helps in applying subsequent graph-based methods for bot detection. The three relationship types are: first-level comment users and post users, second-level comment users and post users, and first-level comment users and second-level comment users. Additionally, the number of user comments can be used as edge weights in the interaction graph structure, providing the model with additional structural information to aid in detection.

\section{D \hspace{0.2cm} Ethics Statement}

To ensure that our work does not cause any potential harm or adverse effects on real-world environments, we affirm that all agent bots and the false information they disseminate are entirely created within this simulation. This dataset includes various information, such as user profiles, posts, comments, and other data valuable for future research. We pledge that this dataset will be used exclusively for scientific research purposes.

\end{document}